\documentclass[aps,prd,eqsecnum,11pt,showpacs,preprintnumbers,nofootinbib]{revtex4}

\def\nbox#1#2{\vcenter{\hrule \hbox{\vrule height#2in
\kern#1in \vrule} \hrule}}
\def\sq{\,\raise.5pt\hbox{$\nbox{.09}{.09}$}\,}
\def\sqb{\,\raise.5pt\hbox{$\overline{\nbox{.09}{.09}}$}\,}

\usepackage{amssymb,amsmath}
\usepackage{graphicx}

\begin{document}

\preprint{\rm LA-UR 07-4898}

\title{Stress Tensor from the Trace Anomaly in Reissner-Nordstr\"{o}m Spacetimes}
\author{Paul R. Anderson}
\affiliation{Department of Physics,\\
Wake Forest University, \\
Winston-Salem, NC 27109 and \\
Racah Institute of Physics, Hebrew University of Jerusalem, Givat
Ram, Jerusalem, 91904, Israel} \;\; \email{anderson@wfu.edu}
\author{Emil Mottola}
\affiliation{Theoretical Division, T-8 \\
Los Alamos National Laboratory \\
Los Alamos, NM 87545}
\email{emil@lanl.gov}
\author{Ruslan Vaulin}
\affiliation{Department of Physics \\
University of Wisconsin-Milwaukee\\
Milwaukee, WI  53211}
\email{vaulin@gravity.phys.uwm.edu}

\begin{abstract}
\vskip .3cm

The effective action associated with the trace anomaly provides a
general algorithm for approximating the expectation value of the
stress tensor of conformal matter fields in arbitrary curved
spacetimes. In static, spherically symmetric spacetimes, the
algorithm involves solving a fourth order linear differential
equation in the radial coordinate $r$ for the two scalar auxiliary fields
appearing in the anomaly action, and its corresponding stress tensor.
By appropriate choice of the homogeneous solutions of the auxiliary
field equations, we show that it is possible to obtain finite stress
tensors on all Reissner-Nordstr\"{o}m event horizons, including
the extreme $Q=M$ case. We compare these finite results to previous
analytic approximation methods, which yield invariably an infinite
stress-energy on charged black hole horizons, as well as with
detailed numerical calculations that indicate the contrary. The
approximation scheme based on the auxiliary field effective action
reproduces all physically allowed behaviors of the quantum stress
tensor, in a variety of quantum states, for fields of any spin, in
the vicinity of the entire family ($0\le Q \le M$) of RN horizons.
\end{abstract}


\maketitle

\vfill\eject

\section{Introduction}

The evaluation of the energy-momentum-stress tensor $T^a_{\ b}$ of quantum
matter in curved spacetimes is important for understanding the possible
backreaction effects of matter on the large scale geometry of spacetime.
Quantitative control of the stress tensor is needed especially in black hole
and cosmological spacetimes with event horizons, where general considerations
indicate that vacuum polarization and particle creation may lead to significant
quantum effects which are cumulative with time. Such secular, macroscopic effects
of quantum matter on the geometry of spacetime provide the intriguing
possibility of observable consequences of quantum gravity at accessible
energy scales, far below the Planck scale.

In the direct method of computing the quantum expectation value of the stress
tensor, the first step is to solve the appropriate matter field equations,
for a complete set of normal modes. With these solutions in hand, the Fock space
of the quantum theory is constructed, a specific ``vacuum" state $\vert\Psi\rangle$
in the Fock space chosen, and the expectation value
$\langle \Psi\vert T^a_{\ b}\vert \Psi \rangle$ evaluated in the selected
state, component by component, as a sum over the normal modes of the field.
Since $T^a_{\ b}$ is a dimension four operator in four spacetime dimensions,
the mode sum for its expectation value is quartically divergent. Hence a
delicate regularization procedure, such as point-splitting must be employed
in order to identify and remove the short distance divergences in the mode
sum, absorbing them into appropriate counterterms up to dimension four
in the gravitational effective action \cite{BirDav}. Only after this
regularization and subtraction procedure is performed can finite results
for the renormalized $\langle T^a_{\ b} \rangle$ of physical interest be
extracted.

Since the wave equation, mode functions, and stress tensors are
different for fields with different spin, this procedure must be
carried out independently for each quantum field of interest.
Likewise, if one wishes to consider different quantum states in the
Fock space, with different boundary conditions on the mode
functions, the calculation must be repeated for each state. Because
of the intricacy of the subtraction procedure, together with the
numerical solution of the mode equations, which is usually required, it is
often difficult to anticipate the general physical features of the
result. Finally, if the geometry is modified, or allowed to
respond dynamically, the entire calculation of $\langle T^a_{\ b}
\rangle$ would have to be repeated for each new geometry and/or at
each new time step. This direct method of calculating $\langle
T^a_{\ b}\rangle$ is thus both time and computation intensive, and
has limited the number of results for the stress tensor in fixed
backgrounds to only a handful of special cases, making the
consideration of dynamical black hole spacetimes varying with time in response
to $\langle T^a_{\ b}\rangle$ prohibitively difficult, even in the
case of exact spherical symmetry.

Because of the difficulties involved in direct evaluation methods,
considerable interest attaches to developing general and reliable algorithms
for approximating $\langle T^a_{\ b}\rangle$ in general curved spacetimes.
An approximate method that can capture the secular, macroscopic quantum
effects on the geometry would be particularly interesting for
applications to both cosmological and black hole spacetimes.

In a previous article a general approximation scheme based on the
effective action and stress tensor obtained from the conformal or
trace anomaly was introduced \cite{MotVau}. Although the effective
action associated with the trace anomaly is not unique, as it is
defined only up to arbitrary conformally invariant terms, a minimal
generally covariant action can be found by direct integration of the
anomaly~\cite{Rie,FraTse}. The logarithmic scaling behavior of the
effective action associated with the anomaly separates it from any
of the other possible local or non-local Weyl invariant terms in the
exact effective action, which do not share this logarithmic scaling
property. The non-local anomaly action may be cast into a local form
in a standard way by the introduction of one or more scalar
auxiliary degree(s) of freedom~\cite{Rie,FraTse,ShaJac,BFS}.
Since there are two distinct cocycles in the non-trivial cohomology
of the Weyl group in four dimensions \cite{MazMot}, the most general
representation of the anomaly action is in terms of two auxiliary scalar
degrees of freedom, each satisfying fourth order linear differential
equations of motion (\ref{EFtraces}). These are two new scalar degrees of
freedom in the low energy effective theory of gravity not present
in the classical Einstein theory. Since the effective action expressed
in terms of the auxiliary fields is a spacetime scalar, variation with
respect to the metric yields  a covariantly conserved stress tensor, which
not only reproduces the trace anomaly but also yields non-trivial tracefree
components as well.

In the auxiliary field approach, computation of the {\it quantum}
expectation value $\langle T^a_{\ b} \rangle$ is reduced to the
solution of linear, {\it classical} equations for the auxiliary
fields, bypassing completely any summation over modes, and the
regularization and renormalization that requires. Different states
of the underlying quantum field(s) are associated with the choice of
specific homogeneous solutions to the linear differential equations
satisfied by the auxiliary fields. This allows for states obeying
different boundary conditions on the horizon to be studied
simultaneously. In addition, matter fields of every spin are treated
in a unified manner, since the auxiliary field equations do not
depend on the spin of the underlying quantum field. The stress
tensor depends on the spin of the fields only through the known spin
dependence of the two numerical coefficients appearing in the
anomaly in eqs. (\ref{bbprime}) below. Finally, the auxiliary field
action and the stress tensor derived from it can be evaluated in principle
in any spacetime, dynamical or not, without respect to special symmetries.
Thus the scalar effective action of the anomaly furnishes a general classical
algorithm for approximating the expectation value of the full stress
tensor of quantum matter of any spin in an arbitrary curved spacetime.

The auxiliary fields are sensitive to macroscopic boundary conditions
and the presence of causal horizons, so it is particularly interesting to
apply the approximation algorithm based on them to spacetimes with event horizons,
where quantum fluctuations are expected to play an important role, and comparison
with existing numerical results is possible. Even when the quantum matter
fields are not strictly massless, their fluctuations and stress tensor in
the vicinity of an event horizon can exhibit conformal behavior \cite{MotVau}.
In particular, quantum states with diverging $\langle T^a_{\ b}\rangle$ on the
Schwarzschild horizon, for which the backreaction on the classical geometry is
significant, have precisely those diverging behaviors prescribed by the stress
tensor derived from the effective action of the anomaly.

In Ref. \cite{BFS} a study of the stress tensor obtained from the anomaly in
Schwarzschild spacetime was undertaken.  In Ref. \cite{MotVau} the general form
of the stress tensor due to the conformal anomaly in an arbitrary spacetime was given
and applied to a few special cases, such as Schwarzschild and de Sitter
spacetimes.  A detailed comparison of the two studies is given
in Section 4.

In the present article we extend and
develop the classical approximation technique of Ref. \cite{MotVau} for the quantum stress
tensor based on the trace anomaly to static, spherically symmetric
spacetimes, focusing specifically on electrically charged
Reissner-Nordstr\"{o}m (RN) black hole spacetimes, and states with
regular stress tensors on the RN horizon. Since direct computations
of the renormalized stress tensor expectation value $\langle T_a^{\
b}\rangle$ have been carried out in both Schwarzschild and
Reissner-Nordstr\"{o}m (RN) spacetimes for free fields of
various spin
\cite{Candelas,Fawcett,Elst1,Elst2,HowCan,How,JenOtt,JenMcOttBR,AnHiSa,AHL,CHOAG},
we will be able to compare the results of the new approximation
scheme to these direct computations of $\langle T^a_{\ b}\rangle$.
Since the {\it exact} effective action of quantum matter generally
contains terms which are not determined by the anomaly, using it to
compute the stress tensor expectation value is certainly an
approximation, which will differ from the direct evaluation of
$\langle T^a_{\ b} \rangle$ in terms of mode sums in general. It is
therefore a non-trivial check of the approximation scheme if all of
the allowed behaviors of the {\it traceless} parts of the exact
$\langle T^a_{\ b} \rangle$ on event horizons can reproduced by the
auxiliary field method based on the trace anomaly.

Several approaches to approximating $\langle T^a_{\ b}\rangle$
have been discussed previously in the literature, developed with important
special cases in mind, such as static geometries with a timelike
Killing field \cite{Page,BroOtt,BOP,FroZel,Zan,AnHiSa,GrAnCa,Pop}. These
approximations are quite successful for regular states in the Schwarzschild
geometry, but fail when compared to the numerical results for $\langle T^a_{\ b} \rangle$
in the charged RN spacetimes. Specifically, the previous approximations
invariably yield a renormalized $\langle T^a_{\ b} \rangle$ which
grows logarithmically without bound 
as the horizon of any charged RN black hole is approached. For the
case of an extreme Reissner-Nordstr\"{om} black hole there is an even
stronger linear divergence.  On the other hand direct numerical evaluation of
$\langle T^a_{\ b}\rangle$ in the Hartle-Hawking-Israel~\cite{HarHaw}
thermal state shows no evidence for any of these divergences \cite{AnHiSa,AHL,CHOAG}.
We review the previous approximation methods and compare them both to
the auxiliary field method and the direct evaluations of $\langle
T^a_{\ b} \rangle$ in Section 4.

Our main purpose in this paper is to show that the auxiliary field
effective action and stress tensor determined by the trace anomaly
leads to a practical semi-analytic approximation technique which
allows for a finite $\langle T^a_{\ b} \rangle$ on the event
horizons of all electrically charged black holes, including the
extreme Reissner-Nordstr\"{o}m (ERN) case of $Q=M$. Although the stress
tensor diverges on the horizon for {\it generic} solutions of the auxiliary
field equations, it is possible to adjust the homogeneous solutions of
these linear equations to remove the divergences.
The ERN case is particularly interesting, since its Hawking
temperature vanishes and its degenerate horizon structure leads to
potentially more severely divergent terms in the stress tensor.
These leading divergent behaviors can be determined analytically by
a power series expansion of the auxiliary fields in the local
vicinity of the horizon, and explicitly cancelled, if desired.

The paper is organized as follows. In the next section we review the
effective action and stress tensor of the trace anomaly in the
auxiliary field form introduced in \cite{MotVau}. In Section 3, we
apply the general approximation algorithm to static, spherically
symmetric spacetimes, reviewing briefly the uncharged Schwarzschild
case, and then extending the analysis to the generic charged $Q<M$
RN cases, and the ERN $Q=M$ case. We determine in each case the
conditions on the series expansion coefficients of the auxiliary
fields necessary for a regular $\langle T^a_{\ b} \rangle$ on the
horizon. In Section 4 we solve the regularity conditions and plot
the results for the simplest solution in each case, comparing them
to previous analytic approximation schemes, and direct numerical
evaluations of $\langle T^a_{\ b} \rangle$. Section 5 contains our
Conclusions, while the Appendix catalogs the complete list of
solutions to the regularity conditions.

\section{Stress Tensor from the Trace Anomaly}

Classical fields satisfying wave equations with zero mass, which are
invariant under conformal transformations of the spacetime metric,
$g_{ab} \rightarrow e^{2\sigma} g_{ab}$ have stress tensors with
zero classical trace, $T^a_{\ a} = 0$. Because the corresponding
quantum theory requires an ultraviolet (UV) regulator, classical
conformal invariance cannot be maintained at the quantum level. The
trace of the stress tensor is generally non-zero when $\hbar \ne 0$,
and any UV regulator which preserves the covariant conservation of
$T^a_{\ b}$, a necessary requirement of any theory respecting
general coordinate invariance, yields an expectation value of the
quantum stress tensor with a non-zero trace: $\langle T^a_{\ a}
\rangle \neq 0$. This conformal or trace anomaly is therefore a
general feature of quantum theory in gravitational fields, on the
same footing as the chiral anomaly in QCD responsible for the
experimentally measured decay of 
the $\pi^0$ meson into two photons~\cite{Adler}.

In four spacetime dimensions the trace anomaly takes the general
form \cite{anom,BirDav},
\begin{equation}
\langle T^a_{\ a} \rangle
= b F + b' \left(E - \frac{2}{3}\sq R\right) + b'' \sq R + \sum_i \beta_iH_i\,.
\label{tranom}
\end{equation}
In Eq. (\ref{tranom}) we employ the notation,
\begin{subequations}
\begin{eqnarray}
&&E \equiv ^*\hskip-.2cmR_{abcd}\,^*\hskip-.1cm R^{abcd} =
R_{abcd}R^{abcd}-4R_{ab}R^{ab} + R^2
\,,\qquad {\rm and} \label{EFdef}\\
&&F \equiv C_{abcd}C^{abcd} = R_{abcd}R^{abcd}
-2 R_{ab}R^{ab}  + \frac{R^2}{3}\,.
\end{eqnarray}
\end{subequations}
\noindent  with $R_{abcd}$ the Riemann curvature tensor,
$^*\hskip-.1cmR_{abcd}= \frac{1}{2}\varepsilon_{abef}R^{ef}_{\ \
cd}$ its dual, and $C_{abcd}$ the Weyl conformal tensor. The
coefficients $b$, $b'$, and $b''$ are dimensionless parameters
proportional to $\hbar$. Additional terms denoted by the sum $\sum_i
\beta_i H_i$ in (\ref{tranom}) may also appear in the general form
of the trace anomaly, if the massless field in question couples to
additional long range gauge fields. Thus in the case of massless
fermions coupled to a background gauge field, the invariant $H
=$tr($F_{ab}F^{ab}$) appears in (\ref{tranom}) with a coefficient
$\beta$ determined by the beta function of the relevant gauge
coupling \cite{AdlColDun}.

The form of (\ref{tranom}) and coefficients $b$ and $b'$ do not
depend on the state in which the expectation value of the stress
tensor is computed. Instead they are determined only by the number of
massless fields and their spin via
\begin{subequations}
\begin{eqnarray}
b &=& \frac{\hbar}{120 (4 \pi)^2}\, (N_S + 6 N_F + 12 N_V)\,,\\
b'&=& -\frac{\hbar}{360 (4 \pi)^2}\, (N_S + 11 N_F + 62 N_V)\,,
\end{eqnarray}
\label{bbprime}
\end{subequations}
\vskip -.3cm
\noindent with $N_S$ the number of spin $0$ fields, $ N_F$ the
number of spin $\frac{1}{2}$ Dirac fields, and $N_V$
the number of spin $1$ fields~\cite{BirDav}.  Henceforth
we shall set $\hbar = 1$, although it should be remembered that
any effect of the anomaly in which the $b$, $b'$ and $\beta_i$
coefficients appear is a one-loop quantum effect.

The trace anomaly determines the conformal variation of the one-loop
effective action of the matter fields in a general curved
background. A covariant, non-local form of this effective action was
first given in Ref.~\cite{Rie}. One consequence of the effective
action due to the anomaly is that the scalar or conformal part of
the metric becomes dynamical, and its fluctuations provide a
mechanism for the screening of the cosmological vacuum energy
\cite{AM,Odint,AMMDE}. The stress tensor, canonical quantization of
the conformal degree of freedom and physical states of the quantum
conformal factor in the absence of the Einstein-Hilbert term were
studied in Ref.~\cite{AMM}.

That the non-local action of the anomaly could be rendered local by
the introduction of scalar auxiliary field(s), was noted in
Refs.~\cite{Rie,FraTse,ShaJac,BFS}. Partial forms of the stress tensor
due to this effective action were given in~\cite{AMM,MazMot,BFS},
with the authors of \cite{BFS} initiating the study of the stress tensor
obtained from the effective action of the anomaly with auxiliary fields
as an approximation scheme in the Ricci flat case of Schwarzschild spacetime.
The general, complete form of the stress tensor in terms of two
auxiliary fields in an arbitrary curved spacetime was given in~\cite{MotVau}.
This auxiliary field effective action is of the form,
\begin{equation}
S_{anom} = b' S^{(E)}_{anom}[g; \varphi] + b S^{(F)}_{anom}[g; \varphi, \psi]\,,
\label{Sanom}
\end{equation}
with
\begin{eqnarray}
&& S^{(E)}_{anom}[g; \varphi] \equiv \frac{1}{2} \int\,d^4x\,\sqrt{-g}\ \left\{
-\left(\sq \varphi\right)^2 + 2\left(R^{ab} - \frac{R}{3}g^{ab}\right)(\nabla_a \varphi)
(\nabla_b \varphi) + \left(E - \frac{2}{3} \sq R\right) \varphi\right\};
{\rm and} \nonumber\\
&& S^{(F)}_{anom}[g; \varphi, \psi] \equiv \,\int\,d^4x\,\sqrt{-g}\
\left\{ -\left(\sq \varphi\right)
\left(\sq \psi\right) + 2\left(R^{ab} - \frac{R}{3}g^{ab}\right)(\nabla_a \varphi)
(\nabla_b \psi)\right.\nonumber\\
&& \hskip 6cm + \left.\frac{1}{2} F \varphi +
\frac{1}{2} \left(E - \frac{2}{3} \sq R\right) \psi \right\}\,,
\label{SEF}
\end{eqnarray}
in terms of the two scalar auxiliary fields $\varphi$ and $\psi$,
corresponding to the two non-trivial cocycles of the Weyl group
in four dimensions \cite{MazMot}.

The effective action (\ref{Sanom})-(\ref{SEF}) is a spacetime scalar integral
over local fields. Hence varying it with respect to the metric yields
two covariantly conserved stress tensors $E_{ab}$ and $F_{ab}$, bilinear
in the scalar auxiliary fields. Explicitly, these are \cite{MotVau}
\begin{eqnarray}
E_{ab} &=&-2\, (\nabla_{(a}\varphi) (\nabla_{b)} \sq \varphi)
+ 2\,\nabla^c \left[(\nabla_c \varphi)(\nabla_a\nabla_b\varphi)\right]
- \frac{2}{3}\, \nabla_a\nabla_b\left[(\nabla_c \varphi)
(\nabla^c\varphi)\right]\nonumber\\
&&
+ \frac{2}{3}\,R_{ab}\, (\nabla_c \varphi)(\nabla^c \varphi)
- 4\, R^c_{\ (a}(\nabla_{b)} \varphi) (\nabla_c \varphi)
+ \frac{2}{3}\,R \,(\nabla_a \varphi) (\nabla_b \varphi)\nonumber\\
&& + \frac{1}{6}\, g_{ab}\, \left\{-3\, (\sq\varphi)^2
+ \sq \left[(\nabla_c\varphi)(\nabla^c\varphi)\right]
+ 2\left( 3R^{cd} - R g^{cd} \right) (\nabla_c \varphi)(\nabla_d
\varphi)\right\}\nonumber\\
&&
\hspace{-1cm} - \frac{2}{3}\, \nabla_a\nabla_b \sq \varphi
- 4\, C_{a\ b}^{\ c\ d}\, \nabla_c \nabla_d \varphi
- 4\, R_{(a}^c \nabla_{b)} \nabla_c\varphi
+ \frac{8}{3}\, R_{ab}\, \sq \varphi
+ \frac{4}{3}\, R\, \nabla_a\nabla_b\varphi \nonumber\\
&&
- \frac{2}{3} \left(\nabla_{(a}R\right) \nabla_{b)}\varphi
+ \frac{1}{3}\, g_{ab}\, \left\{ 2\, \sq^2 \varphi
+ 6\,R^{cd} \,\nabla_c\nabla_d\varphi
- 4\, R\, \sq \varphi
+ (\nabla^c R)\nabla_c\varphi\right\}\,,
\label{Eab}
\end{eqnarray}
and
\begin{eqnarray}
&& F_{ab} = -2\, (\nabla_{(a}\varphi) (\nabla_{b)} \sq \psi)
-2\, (\nabla_{(a}\psi) (\nabla_{b)} \sq \varphi)
+ 2\,\nabla^c \left[(\nabla_c \varphi)(\nabla_a\nabla_b\psi)
 + (\nabla_c \psi)(\nabla_a\nabla_b\varphi)\right]
\nonumber\\
&&
\hspace{.5cm} - \frac{4}{3}\, \nabla_a\nabla_b\left[(\nabla_c \varphi)
(\nabla^c\psi)\right]
+ \frac{4}{3}\,R_{ab}\, (\nabla_c \varphi)(\nabla^c \psi)
- 4\, R^c_{\ (a}\left[(\nabla_{b)} \varphi) (\nabla_c \psi)
+ (\nabla_{b)} \psi) (\nabla_c \varphi)\right]\nonumber\\
&&
\hspace{1.5cm} + \frac{4}{3}\,R \,(\nabla_{(a} \varphi) (\nabla_{b)} \psi)
+ \frac{1}{3}\, g_{ab}\, \Big\{-3\, (\sq\varphi)(\sq\psi)
+ \sq \left[(\nabla_c\varphi)(\nabla^c\psi)\right]
\nonumber\\
&&
\hspace{1.5cm} \left. + 2\, \left( 3R^{cd} - R g^{cd} \right) (\nabla_c
\varphi)(\nabla_d \psi)\right\}- 4\, \nabla_c\nabla_d\left( C_{(a\ b)}^{\ \ c\ \ d}
\varphi \right)  - 2\, C_{a\ b}^{\ c\ d} R_{cd} \varphi \nonumber\\
&&
\hspace{.5cm} - \frac{2}{3}\, \nabla_a\nabla_b \sq \psi
- 4\, C_{a\ b}^{\ c\ d}\, \nabla_c \nabla_d \psi
- 4\, R_{(a}^c (\nabla_{b)} \nabla_c\psi)
+ \frac{8}{3}\, R_{ab}\, \sq \psi
+ \frac{4}{3}\, R\, \nabla_a\nabla_b\psi \nonumber\\
&&
\hspace{1cm} - \frac{2}{3} \left(\nabla_{(a}R\right) \nabla_{b)}\psi +
\frac{1}{3}\, g_{ab}\,  \left\{ 2\, \sq^2 \psi + 6\,R^{cd} \,\nabla_c\nabla_d\psi
- 4\, R\, \sq \psi + (\nabla^c R)(\nabla_c\psi)\right\}\,.
\label{Fab}
\end{eqnarray}
These tensors have the local geometrical traces,
\begin{subequations}
\begin{eqnarray}
E^a_{\ a} &=& 2 \Delta_4 \varphi = E - \frac{2}{3} \sq R\,,\\
F^a_{\ a} &=& 2 \Delta_4 \psi = F = C_{abcd}C^{abcd}\,,
\label{Ftr}
\end{eqnarray}
\label{EFtraces}
\end{subequations}
\vskip -.7cm
\noindent
where the latter half of these two equations follow from the independent
Euler-Lagrange variation of (\ref{Sanom})-(\ref{SEF}) with respect to the
two auxiliary scalar degrees of freedom, $\varphi$ and $\psi$.

The fourth order scalar differential operator
appearing in these expressions is \cite{Rie,AMM,MotVau}
\begin{equation}
\Delta_4 \equiv \sq^2 + 2 R^{ab}\nabla_a\nabla_b - \frac{2}{3} R \sq +
\frac{1}{3} (\nabla^a R)\nabla_a =
\nabla_a\left(\nabla^a\nabla^b + 2 R^{ab} - \frac{2}{3} R g^{ab}\right)
\nabla_b\,.
\label{Deldef}
\end{equation}
By solving the fourth order linear equations (\ref{EFtraces})
determined by this $\Delta_4$ for the two auxiliary fields, $\varphi$
and $\psi$, and substituting the results into the stress tensors
(\ref{Eab}) and (\ref{Fab}) we obtain a general approximation
algorithm for $\langle T^{\mu}_{\ \nu} \rangle$ for conformal matter fields of
any spin in an arbitrary curved spacetime. That is,
\begin{equation}
\langle T^{\mu}_{\ \nu} \rangle \simeq T^{\mu}_{\ \nu}[\varphi, \psi] =
b' E^{\mu}_{\ \nu} + b F^{\mu}_{\ \nu}\,.
\label{Tanom}
\end{equation}
is an approximation to the exact stress tensor
expectation value. Since the dependence of the stress tensor
$T^{\mu}_{\ \nu}[\varphi, \psi]$ on the spin of the underlying quantum
matter fields arises purely through the numerical coefficients $b$ and $b'$
through (\ref{bbprime}), and $\beta_i$, there are no new equations to be
solved for quantum fields of different spin.

The freedom to add homogeneous solutions of (\ref{EFtraces}) to any
given inhomogeneous solution allows the tracefree components of the
stress tensor (\ref{Tanom}) to be changed without altering its trace.
This corresponds to the freedom to change the boundary conditions
and the state of the underlying quantum field theory without changing
its state independent trace anomaly. As shown in \cite{MotVau} the
auxiliary fields and traceless terms in the stress tensor (\ref{Tanom})
generally diverge on event horizons, which provides a coordinate invariant
meaning to large quantum backreaction effects on horizons. These state
dependent divergences can be analyzed and removed by specifying
boundary conditions for the auxiliary fields on the horizon. One then
has an approximation scheme for the expectation value $\langle T^{\mu}_{\ \nu}
\rangle$ in regular states as well.

In order to characterize the nature of the approximation (\ref{Tanom}),
we recall the general decomposition of the exact quantum effective
action into three parts,
\begin{equation}
S_{exact} = S_{local} + S_{inv} + S_{anom}\,,
\label{Sexact}
\end{equation}
according to its transformation properties under global Weyl
transformations \cite{MazMot}. The local action $S_{local}$ can be
expressed purely in terms of local contractions of the Riemann
curvature tensor and its derivatives. In addition to the classical
Einstein-Hilbert action $S_{local}$ consists of an infinite series
of higher dimension curvature invariants multiplied by increasing
powers of an inverse mass scale. These terms, consistent with a
general effective field theory analysis of gravity, give higher
order geometric contributions to the stress tensor, which remain
bounded and small for small curvatures. The two remaining terms in
(\ref{Sexact}) are generally non-local. Any non-local terms
involving a non-zero {\it fixed} mass parameter can be expanded in a
series of higher derivative local terms multiplied by powers of the
inverse mass and regrouped into $S_{local}$. Hence we need consider
only those non-local terms which are {\it not} associated with any
fixed mass or length scale in the remaining terms of $S_{exact}$.
These must be either strictly Weyl invariant, denoted by $S_{inv}$
in (\ref{Sexact}), or break local Weyl invariance, yet without
introducing any explicit mass or length scale. Up to possible
surface terms, these are just the geometric terms required by the
trace anomaly (\ref{tranom}). The associated terms in $S_{anom}$
scale logarithmically with distance or energy, and are composed of
the two distinct cocycles of the Weyl group, given by
(\ref{Sanom}).

When the background spacetime is conformally flat or approximately
so, the Weyl invariant action $S_{inv}$ may be neglected, since it
vanishes in the conformally related flat spacetime, in the usual
Poincar\'e invariant vacuum state. In that case we expect $S_{anom}$
to become a good approximation to the non-local terms in the exact
effective action (\ref{Sexact}), and the corresponding stress tensor
(\ref{Tanom}) to become a good approximation to the exact quantum
stress tensor, up to well-known local terms. Thus the approximation
(\ref{Tanom}) amounts to the neglect of $S_{inv}$, or more precisely,
those parts of $S_{inv}$ which cannot be expressed in terms of local
terms or parameterized by homogeneous solutions to the auxiliary field
equations (\ref{EFtraces}).

Because of the conformal behavior of fields near an event horizon,
where the effects of mass terms become subdominant, one might expect
the leading behavior of the stress tensor (\ref{Tanom})
also to match that of the exact $\langle T^a_{\ b} \rangle$ in the
vicinity of the horizon. In \cite{MotVau} we tested this hypothesis
in Schwarzschild and de Sitter spacetimes, finding that the freedom
to choose homogeneous solutions to the auxiliary field equations
(\ref{EFtraces}) allows for all possible allowed behaviors of the
exact stress tensor near the horizon. Indeed for states with
diverging stress tensor on the horizon, the anomalous stress tensor
(\ref{Tanom}) gives the correct leading and subleading behaviors of
the exact $\langle T^a_{\ b}\rangle$.

When attention is restricted to states with regular behavior on the
horizon, which from the point of view of conformal invariance are
{\it subleading} with respect to the divergent terms, then the
stress tensor (\ref{Tanom}) can often be adjusted to give the exact finite
value of $\langle T^a_{\ b}\rangle$ on the horizon as well. However,
the global fit of (\ref{Tanom}) is in only fair quantitative
agreement with the numerically computed expectation value $\langle
T^a_{\ b}\rangle$ far from the horizon for the Hartle-Hawking-Israel state
in Schwarzschild spacetime. In regular states the neglected terms in
$S_{exact}$, which remain bounded at the horizon, are comparable in
magnitude to subleading terms of $S_{anom}$. Hence neglect of
$S_{inv}$ is expected to yield a poorer global approximation to the
stress tensor of such regular states even when it is possible
to match the exact behavior of $\langle T^a_{\ b}\rangle$ at the
horizon.

The Schwarzschild and de Sitter cases considered in \cite{MotVau}
are special in that the auxiliary field equations (\ref{EFtraces})
may be solved analytically. However, the approximation (\ref{Tanom})
does not require this, and in the following we extend the auxiliary
field method to general static, spherically symmetric geometries,
focusing specifically on
the RN family of charged black holes. These
black hole spacetimes provide an interesting testbed for the
auxiliary field stress tensor, whose qualitative and quantitative
features may be compared with both previous approximation methods
and direct numerical evaluation of $\langle T^a_{\ b}\rangle$.

\section{Regular Stress Tensors in Reissner-Nordstr\"{o}m Spacetimes}

The effective action and stress tensor of the auxiliary fields
(\ref{Tanom}) is defined in any spacetime, regardless of
special symmetries. However it becomes particularly useful as
a method of approximating the expectation value of the quantum
stress tensor in spacetimes with a high degree of symmetry,
such as spherical symmetry. The line element for a general static,
spherically symmetric spacetime can be expressed in terms of two
functions of the radius in the form,
\begin{equation}
ds^2 = -f(r)\,dt^2 + \frac{dr^2}{h(r)} + r^2 d\Omega^2\,.
\label{static}
\end{equation}
Assuming that the state in which we evaluate the the stress tensor
is also spherically symmetric and stationary in time, we may make a
static, spherically symmetric ansatz for the auxiliary fields as
well, {\it i.e.}
\begin{subequations}
\begin{eqnarray}
\varphi &=& \varphi(r)\\
\psi &=& \psi(r)\,,
\end{eqnarray}
\label{ppsistatic}
\end{subequations}

\noindent
so that the equations (\ref{EFtraces}) become ordinary
differential equations in $r$. In some cases it is possible also to
add terms with linear time dependence to the auxiliary fields, {\it
i.e.} $\varphi = \varphi(r) + \alpha t$ and and $\psi = \psi(r) +
\alpha' t$ in order to allow for the possibility of non-vanishing
$T^r_{\ t}$ flux components which are also independent of the
Killing time $t$. Higher powers of $t$ or more complicated time
dependence in the auxiliary fields lead to non-stationary stress tensors.

For a general static, spherically symmetric spacetime, with the fields
in a spherically symmetric quantum state, the stress tensor is given by
its three independent diagonal components,
\begin{eqnarray}
T_{\ t}^t & = & -\rho(r) \\
T_{\ r}^r & = & p(r) \\
T_{\ \theta}^{\theta} & = & p_{\bot}(r) \,,
\end{eqnarray}
together with a possible non-zero off-diagonal flux component $T^r_{\ t}$.
These components obey the covariant conservation conditions,
\begin{equation}
\label{cons}
\nabla_a T_{\ r}^a = \frac{d p}{d r}+\frac{1}{2f}\frac{df}{dr}\, (p+\rho)
+ \frac{2}{r}\, (p-p_{\bot})=0
\end{equation}
and
\begin{equation}
\nabla_a T_{\ t}^a = \frac{1}{\sqrt{-g}} \frac{d}{dr}\,
(\sqrt{-g}\, T_{\ t}^r)  = 0\,.
\label{Ttr}
\end{equation}
Eq.\ (\ref{Ttr}) can be trivially integrated to obtain
\begin{equation}
T^r_{\ t} = -\frac{L}{4\pi r^2}{\sqrt\frac{h}{f}}
\label{Ttrgen}
\end{equation}
with the integration constant $L$ the
luminosity of a localized source.

The approximation of (\ref{Tanom}) can be applied to arbitrary spacetimes,
spherically symmetric or not, with or without an horizon. For definiteness
we restrict our attention in this paper to the family of
Reissner-Nordstr\"{o}m spacetimes with the equal metric functions,
\begin{equation}
f(r) = h(r) = 1 - \frac{2M}{r} + \frac{Q^2}{r^2}\,,
\label{fhRN}
\end{equation}
depending upon $M$, the mass and $Q$, the electric charge.
We then can distinguish three cases:
\renewcommand{\theenumi}{(\roman{enumi})}
\begin{enumerate}
\item  $Q=0$, Schwarzschild spacetime;
\item $0<Q<M$, generic Reissner-Nordstr\"{o}m (RN) spacetime;
\item $Q=M$, extreme Reissner-Nordstr\"{o}m (ERN) spacetime.
\end{enumerate}
We shall discuss each of these cases in detail separately.

Since (\ref{fhRN}) is quadratic in $1/r$ there are two values of $r$ at which
$f(r)$ vanishes.  These are
\begin{equation}
r_{\pm} = M \pm \sqrt{M^2-Q^2}\,.
\end{equation}
When $Q=0$, $r_+ = 2M$ is the usual Schwarzschild horizon. In the
ERN case of $Q=M$, the two values $r_\pm$ coincide and the horizon
becomes degenerate. The character of the spacetime, the solutions to
the auxiliary field equations and the corresponding stress tensors
derived from them are quite different in each of the three cases.
Henceforth we reserve the designation RN for the generic, charged
black hole solution of case (ii).

For conformal field theories, the trace of the stress
tensor is given purely by the trace anomaly. This provides us with
another relation for the diagonal components of the tensor, namely
\begin{equation}
-\rho + p + 2p_{\bot} \equiv T = b' E + b F
\label{radtr}
\end{equation}
Furthermore, defining
\begin{equation}
\Theta(r) \equiv p_{\bot}- \frac{T}{4}
\label{defth}
\end{equation}
and integrating (\ref{cons}) with (\ref{radtr}) gives~\cite{ChrFul}
\begin{equation}
p(r)=\frac{1}{r^2f}\int^r_{r_+} dr\left(2rf-r^2f'\right)\Theta +
\frac{1}{4r^2f}\int^r_{r_+}dr\, \frac{d(r^2f)}{dr}\, T +\frac{C-L}{4\pi r^2f},
\label{intp}
\end{equation}
where $C\,-\,L$ denotes an overall constant of integration. The only
unknown function in this expression is $\Theta(r)$. Thus it is
sufficient to compute $\Theta (r)$ from the auxiliary field stress
tensor to determine all the diagonal components of the stress
tensor, $p(r)$, $p_{\bot}(r)$ and $\rho(r)$ from equations
(\ref{intp}), (\ref{defth}) and (\ref{radtr}) respectively. From 
Eq. (\ref{Ttrgen}) with $h(r)=f(r)$, the off
diagonal flux component is
\begin{equation}
T^r_{\ t} = -\frac{L}{4\pi r^2} \;.
\label{flux}
\end{equation}

Since the curvature invariants in the trace (\ref{radtr}) remain finite
as the horizon at $r=r_+$ is approached, it is clear from (\ref{defth})
that $p_{\bot}$ is finite on the horizon, provided $\Theta(r)$
remains finite there. Moreover if $f(r)$ possesses
an isolated simple zero at $r=r_+$, vanishing linearly in
\begin{equation}
s=(r-r_+)/r_+\,.
\label{sdef}
\end{equation}
as $r\rightarrow r_+$, so that the event horizon is non-degenerate,
then (\ref{intp}) shows that a divergent term in the other
components of the stress tensor can arise only if $C-L \neq 0$.
Hence $\Theta(r)$ remaining finite as $r\rightarrow r_+$, and $C=L$,
are necessary and sufficient conditions for finiteness of all
components of $\langle T^a_{\ b}\rangle$ in the static Killing
frame of (\ref{static}) for the non-degenerate Reissner-Nordstr\"{o}m
horizons ($0 \le Q < M$).

Finiteness of the stress tensor in the frame of a freely falling
observer requires also that~\cite{ChrFul}
\begin{equation}
\label{Tuu}
\frac{|T_{uu}|}{f^2} =\frac{1}{4f^2}\Big\vert (\rho+p)f +\frac{L}{2\pi r^2}
\Big\vert  < \infty\,.
\end{equation}
If $\Theta(r)$ is regular at $r=r_+$ and can be expanded in a Taylor
series in $s$ near $s=0$, then it is easy to see that the condition
(\ref{Tuu}) is satisfied automatically if $C=0$. In the case of the
past horizon one should demand finiteness of $\vert T_{vv}\vert
/f^2$ instead of (\ref{Tuu}). This forces the integration constant
to be $C=2L$ instead. Thus regularity on both the past and future
horizon requires $\Theta (r_+) < \infty$ and $C=L=0$.  In this case
the regularity condition (\ref{Tuu}) is equivalent to the
condition
 \begin{equation}
 |T^{uu}| = \frac{1}{f} |\rho + p| <\infty \;.  \label{Tuucondition}
\end{equation}
In some cases, {\it i.e.} those with logarithmic terms in the
auxiliary fields, $\Theta(r)$ is not analytic at $r=r_+$ and cannot
be expanded in a Taylor series there. In those cases, the finiteness
condition (\ref{Tuu}) will give additional conditions on the
behavior of the auxiliary fields on the horizon.

Before restricting our attention to the static, completely regular
class of stress tensors, let us emphasize that this is not the
generic case, as a different choice of the free integration
constants $C$ and $L$ would lead to different physical behavior on
the horizon. It is well known that the coordinate singularity of the
metric (\ref{static}) at the horizon $r=r_+$ where $f=h$ vanishes
may be removed by a (singular) coordinate transformation, producing
the complex analytic extension of the Reissner-Nordstr\"{o}m geometry
\cite{HawEll}. Although any regular
transformation of coordinates is allowed by the Equivalence
Principle, and cannot lead to physical effects, {\it singular}
coordinate transformations, like singular gauge transformations in
gauge theory, must be treated with some care. New topological
configurations such as monopoles or vortices are associated with
such singular gauge transformations. Thus, although the
complex analytic extension of a black hole spacetime may seem quite
natural mathematically, analytic continuation actually involves a
physical assumption, namely that there are no stress tensor sources
to the Einstein equations localized on or near the horizon. Because
of the hyperbolic nature of Einstein's equations such stress sources
with effects transmitted along a null surface are perfectly
allowable, even classically. When the expectation value of the
stress tensor of quantum fields is considered, with its sensitivity
to the wavelike, non-local coherence effects of quantum matter, the
assumption of analyticity on the horizon is not at all automatic,
and is not required by any general principles of quantum theory.

The effective action of the conformal anomaly, and its associated
auxiliary fields indicate that non-regularity of the stress tensor
on the horizon is to be expected in the generic case as well
\cite{MotVau}. A tuning of the integration constants of the
solutions of the linear equations (\ref{EFtraces}) for $\varphi$ and
$\psi$ is necessary to prevent $\Theta (r)$ from diverging as
$r\rightarrow r_+$, with the generic behavior of all the stress
tensor components near the horizon being proportional to $f^{-2}$ as
$f \rightarrow 0$ in the Schwarzschild case. Since the auxiliary
fields are spacetime scalars, it is clear that this behavior is in
no contradiction to the Equivalence Principle. In fact, the
divergences have a perfectly coordinate invariant origin in terms of
the homogeneous solutions to the equations (\ref{EFtraces})
\begin{equation}
\varphi_h = \psi_h = \ln \left(-K^aK_a\right) = \ln f(r)
\label{Kill}
\end{equation}
where $K = \frac{\partial}{\partial t}$ is the Killing field of
the static Schwarzschild or Reissner-Nordstr\"{o}m geometries, timelike
for $r>r_+$. This defines the rest frame of the configuration,
which is independent of coordinate redefinitions. The allowed divergences
in the stress tensor on the horizon as $f(r) \rightarrow 0$ and $\varphi, \psi \rightarrow
\pm \infty$ are related therefore to the behavior of this Killing invariant
of the global geometry becoming null. The state of the quantum matter fields,
specified on a complete Cauchy surface of the spacetime is necessarily
defined in a non-local way, and hence expectation values of $T^a_{\ b}$
can be sensitive to the divergences of (\ref{Kill}) at this null surface,
notwithstanding the finiteness of the local Riemann curvature at $r=r_+$.

By choosing suitably restricted solutions of the auxiliary field
equations (\ref{EFtraces}) it is possible to cancel the $\ln f$
behavior in $\varphi$ and $\psi$, thereby guaranteeing that
$\Theta (r_+)$ is finite. When the electric charge satisfies $0<Q<M$, {\it
i.e.} excluding the uncharged Schwarzschild and maximally charged
ERN cases, the stress tensor will have subleading $\ln s$ and $\ln^2
s$ divergences as $s\rightarrow 0$ as well. Then $\Theta (r)$ is not
expandable as a Taylor series around $r=r_+$ and the logarithmic
terms do not drop out entirely from the condition (\ref{Tuu}). This
leads to additional conditions on the coefficients of the expansion
around $s=0$ for a fully regular stress tensor.

\subsection{Schwarzschild Spacetime}

In the Schwarzschild case, $f(r) = 1 - 2M/r$, and the fourth order
linear equations (\ref{EFtraces}) can be integrated explicitly for
auxiliary fields which are functions only of $r$. The result is
\cite{BFS,MotVau}
\begin{eqnarray}
\frac{d\varphi}{dr}\Big\vert_{_S}
&=&\frac{q-2}{6M}\,\left(\frac{r}{2M} + 1 + \frac{2M}{r}\right)
\ln \left(1 - \frac{2M}{r}\right) - \frac{q}{6r}\left[\frac{4M}{r-2M}
\ln \left(\frac{r}{2M}\right) + \frac{r}{2M} + 3 \right] \nonumber\\
&& \qquad  - \frac{1}{3M} - \frac{1}{r} + \frac{2M c_{_H}}{r (r-2M)}
+ \frac{c_\infty}{2M}\, \left(\frac{r}{2M} + 1 + \frac{2M}{r}\right)
\label{phipS}
\end{eqnarray}
in terms of the three dimensionless constants of integration,
$c_{_H}$, $c_\infty$, and $q$. A fourth integration constant would
be introduced by integrating (\ref{phipS}) once further, but as
the stress tensor in the Schwarzschild case depends only upon
derivatives of $\varphi$, a constant shift in $\varphi$ plays
no role in this case.

The role of the three integration constants appearing in (\ref{phipS})
is best exposed by examining the limits,
\begin{subequations}
\begin{eqnarray}
&&\hskip-1.5cm \frac{d\varphi}{dr}\Big\vert_{_S} \rightarrow \frac{c_{_H}}{r-2M}
+ \frac{q-2}{2M} \,\ln\left(\frac{r}{2M} - 1\right) - \frac{1}{2M}
\left(3 c_\infty - c_{_H} - q - \frac{5}{3}\right) + \dots,\qquad r
\rightarrow 2M,
\label{Schlima}\\
&&\hskip-1.5cm \frac{d\varphi}{dr}\Big\vert_{_S} \rightarrow
\frac{c_\infty r}{4M^2} +
\frac{2c_\infty - q}{4M} + \frac{c_\infty}{r} - \frac{2M}{3r^2}\,q\,
\ln \left(\frac{r}{2M}\right) + \frac{2M}{r^2}\left[c_{_H} - \frac{7}{18}
(q - 2)\right] + \dots,\,r \rightarrow \infty .
\label{Schlimb}
\end{eqnarray}
\end{subequations}
Hence $c_{_H}$ controls the leading behavior as $r$ approaches the
horizon, corresponding to the homogeneous solution of (\ref{Kill}).
It is this leading behavior that gives rise to the generic $f^{-2}$
behavior of the stress tensor as $r\rightarrow r_+$.
The second integration constant $c_\infty$ controls the leading behavior
of $\varphi(r)$ as $r\rightarrow \infty$, which is the same as in flat
space. Non-zero values of $c_{\infty}$ correspond to non-trivial
boundary conditions at some large but finite volume, such as may
be appropriate in the Casimir effect, or if the black hole is enclosed
in a box. The constant $q$ is the topological charge of the auxiliary
field configuration, associated with the conserved current generated by
the Noether symmetry of the effective action (\ref{Sanom}),
$\psi \rightarrow \psi + const.$ \cite{MotVau}. It is responsible for
the $\ln r$ terms in (\ref{Schlimb}) and the corresponding stress
tensor (\ref{Eab}).

To the general spherically symmetric static solution (\ref{phipS}) we
may add also a term linear in $t$, {\it i.e.} we may replace $\varphi(r)$ by
\begin{equation}
\varphi (r,t) = \varphi (r) + \frac{\eta}{2M}\,t\,,
\label{tSch}
\end{equation}
with $\eta$ an additional free constant of integration.
Linear time dependence in the auxiliary fields is the only allowed
time dependence that leads to a time-independent stress-energy,
and this only in the Ricci flat Schwarzschild case.
The solution for $\psi = \psi(r,t)$ is of the same form as
(\ref{phipS}) and (\ref{tSch}) with four new integration constants,
$d_{_H}, d_\infty$, $q'$ and $\eta'$ replacing $c_{_H}, c_\infty$, $q$,
and $\eta$ in $\varphi(r,t)$. Adding terms with any higher powers of
$t$ or more complicated $t$ dependence produces a time dependent
stress-energy tensor. The stress tensor (\ref{Tanom})
does not depend on either a constant $\varphi_0$ or $\psi_0$.

The stress-energy diverges on the horizon in an entire family of states
for generic values of the eight auxiliary field parameters
$(c_{_H}, q, c_{\infty}, \eta; d_{_H}, q', d_{\infty}, \eta')$. Hence in
the general allowed parameter space of spherically symmetric macroscopic
states, horizon divergences of the stress-energy are quite generic, and
not restricted to the Boulware state \cite{Boul}. In addition to the leading
$s^{-2}$ behavior, there are subleading $s^{-1}$, $\ln^2 s$ and $\ln s$
divergences in general. It turns out that only three of these
four are independent, and the three conditions,
\begin{subequations}
\begin{eqnarray}
&& \label{Scond1}-(b'c_H+2bd_H)c_H + \eta(b'\eta + 2b\eta') = 0 \qquad
\left(s^{-2} \ {\rm in} \ T^{\theta}_{\ \theta}\right)\\
&& \label{Scond2} (q-2)[b'(q-2)+2b(q'-2)]=0 \qquad
\left(\ln^2 s \ {\rm in}\ T^{\theta}_{\ \theta}\right)\\
&&\label{Scond3} b[(q-2)(18d_{\infty}-30d_H-40)+
(q'-2)(18c_{\infty}-30c_H-40)]+\nonumber\\
&& \qquad + b'(q-2)(18c_{\infty}-30c_H-40)=0 \qquad
\left(\ln s \ {\rm in}\ T^{\theta}_{\ \theta}\right)
\end{eqnarray}
\label{ScondA}
\end{subequations}
\vskip -1cm
\noindent are all that are required to remove the divergent behaviors
in $\Theta$, indicated in parentheses as $s\rightarrow 0$, including
a possible $s^{-1}$ divergence at the horizon. The first of these conditions,
(\ref{Scond1}) corrects a sign error in Eq. (5.14b) of Ref. \cite{MotVau}
(where the notations $p,p'$ were used in place of the present $\eta,\eta'$
for the parameters of the linear time dependent terms in $\varphi, \psi$
respectively).

The regularity condition (\ref{Tuu}) on $T^{uu}$ gives in
addition,
\begin{subequations}
\begin{eqnarray}
&& \label{lum}
b\left(\frac{2-q'}{3}+2c_H+2d_H\right)
+ b'\left(\frac{2-q}{3}+2 d_H\right) = b'\eta q + b(q\eta' + q'\eta)
 = \frac{LM^2}{\pi} \\
&& \hskip 2cm \left(s^{-2} \ {\rm in} \ T^{uu}\right)\nonumber\\
&& b [c_H(q'-2)+d_H(q-2)]+b'd_H(q-2)=0  \ \left(\ln s \ {\rm
in} \ T^{uu}\right) 
\label{loguu}
\end{eqnarray}
\label{ScondB}
\end{subequations}
\vskip -1cm \noindent If we are interested in strictly static,
regular states with $C=L=0$ in (\ref{intp}) and (\ref{flux}), then
we obtain an additional condition, namely the quantity in
(\ref{lum}) proportional to the luminosity must vanish. The
condition (\ref{loguu}) eliminates a possible subleading logarithmic
divergence in $T_{uu}/f^2$ on the horizon. It is clear that the choice
$q=q'=2$ satisfies this condition as well as (\ref{Scond2}) and
(\ref{Scond3}). This illustrates the general property that when
logarithmic terms are taken to vanish on the horizon, the conditions that
$\Theta(r_+)$ be finite on the horizon and $C=L=0$ are sufficient to 
yield a fully regular stress tensor on the horizon, and the number 
of independent conditions is reduced.

\subsection{Generic Charged RN Spacetimes}

The generic $Q>0$ charged Reissner-Nordstr\"{o}m spacetimes are not
Ricci flat and an analytic solution of the fourth order equations
(\ref{EFtraces}) in closed form no longer appears possible. In
addition, another difference from the Schwarzschild case, stemming from the
same non-vanishing of the Ricci tensor when $Q>0$, is that the
linear time dependence (\ref{tSch}) in the auxiliary field $\varphi$
now produces time dependence in the stress tensor (\ref{Fab}), and
hence is disallowed for a static state. Linear time dependence in $\psi$
is allowed, leading however to a non-zero flux. Hence for strictly
static states we must set $\eta=\eta' =0$. However, the non-Ricci
flatness means also that a possible constant term in $\varphi$ which
drops out of the stress tensor (\ref{Tanom}) when $R_{ab} = 0$ now
survives as a non-trivial free parameter in this case. Hence we have
seven remaining integration constants in all for spherically symmetric 
static auxiliary fields (\ref{ppsistatic}) in the charged RN case.

Since the fourth order differential operator, $\Delta_4$ involves a
total derivative, {\it cf.} Eq. (\ref{Deldef}), equations
(\ref{EFtraces}) can be integrated once, to obtain the
{\it second} order equation for $\varphi^{\prime} (r)$,
\begin{equation}
fr^2 \frac{d}{dr} \left[\frac{1}{r^2} \frac{d}{dr} (fr^2 \varphi')\right]
-2Q^2 \frac{f}{r^2}\, \varphi' = e_0 + \frac{1}{2} \int^r_{r_+} \,r^2\,dr\,
E \,,
\label{sphphi}
\end{equation}
where $e_0$ is an integration constant. Here we have used the facts that
\begin{equation}
R^r_{\ r} = -\frac{Q^2}{r^4}
\end{equation}
and $R = 0$ in a general RN spacetime.

A power series representation of the general solution of the second
order equation (\ref{sphphi}) for $\varphi'$ in powers of $s = (r-r_+)/r_+$
is easily derived. Since the right side of (\ref{sphphi}) is regular at
$s=0$, it can be expressed as a Taylor series in the form,
\begin{equation}
e_0 + \frac{1}{2} \int^r_{r_+}\,r^2\,dr\, E
= \sum_{n=0} e_ns^n\,.
\label{Eseries}
\end{equation}
To find the leading behavior of $\varphi'$ near $s=0$ let
$\varphi' \sim s^\gamma$ for some $\gamma$, and thereby
obtain from (\ref{sphphi}) and (\ref{Eseries}),
\begin{eqnarray}
&&\epsilon^2 \,\gamma (\gamma + 1)\, s^\gamma + 2\,\epsilon\,
\gamma (\gamma + 2)\,\frac {Q^2}{r_+^2}\, s^{\gamma + 1}\nonumber\\
&& + \left\{ (\gamma + 1)(\gamma + 2) - \frac{2Q^2}{r_+^2}
+ 2\epsilon\,\left[-(\gamma+1)^2 + \frac{2Q^2}{r_+^2}\right] \right\}
\,s^{\gamma +2} + {\cal O} (s^{\gamma +3}) \nonumber\\
&& \qquad =\sum_{n=0} e_ns^n\, ,
\label{leadpower}
\end{eqnarray}
where
\begin{equation}
\epsilon \equiv \frac{r_+-r_-}{r_+} = \frac{2 \sqrt{M^2-Q^2}}{M+ \sqrt{M^2 - Q^2}}
\label{epsdef}
\end{equation}
in the general case. From (\ref{leadpower}) we observe that $\gamma = -1$
gives the most singular behavior allowed for $\varphi'$ as $s\rightarrow 0$ for
general $e_0 \ne 0$, and $r_+ > r_-$. Note that $\varphi' \sim s^{-1}$
agrees also with the leading behavior in (\ref{Schlima}) for the exact solution
in the uncharged Schwarzschild case.

Because the singular $s^{-1}$ behavior is allowed for $\varphi'$
for all $0 \le Q < M$, $\varphi$ has at most logarithmically singular
behavior near the RN horizon, and its general series expansion
is of the form,
\begin{equation}
\varphi(r) = \sum_{n=0}a_ns^n + \sum_{n=0} \ell_ns^n\,\ln s \;.
\label{phiseries}
\end{equation}
Note that the general logarithmic behavior near $s=0$ is that expected from
(\ref{Kill}) on geometrical grounds, since $e^{-\varphi_h}$ with $\varphi_h =
\ln s$ is the conformal transformation needed to bring the RN metric
to the ultrastatic optical metric, and remove the singularity caused
by the Killing field changing character from timelike to null.

Substitution of the expansion (\ref{phiseries}) into the differential equation
(\ref{sphphi}) gives recursion relations for the coefficients $\{a_n, \ell_n\}$.
The set of four coefficients, $(a_0,a_1,\ell_0,\ell_1)$ are free integration constants
for $\varphi$, parameterizing the general solution of the fourth order equation, with
all higher order $a_{n>1}$ and $b_{n>1}$ determined by the recursion relations
in terms of these four integration constants. In the Schwarzschild limit,
$\ell_0 \rightarrow c_H, \ell_1 \rightarrow q-2, a_1 \rightarrow c_H + q
+ \frac{5}{3} - 3 c_{\infty}$ respectively. The constant $e_0$ is also
determined in terms of $\ell_0$ by (\ref{sphphi}) or (\ref{leadpower}) to be
\begin{equation}
e_0 = -2 \epsilon\,\ell_0 \frac{Q^2}{r_+^2}\,.
\end{equation}
In like manner substitution of the series expansion,
\begin{equation}
\psi(r) = \sum_{n=0}c_ns^n + \sum_{n=0}\lambda_ns^n\, \ln s \,,
\label{psiseries}
\end{equation}
into the equation for $\psi$ shows that $(c_0,c_1,\lambda_0,\lambda_1)$ are the
four free integration constants parameterizing the general solution of (\ref{Ftr}).
However, since a constant shift in $\psi$ is an exact symmetry of the
action (\ref{Sanom}), $c_0$ does not appear in the stress tensor (\ref{Tanom}),
and we are left with the seven effective free integration constants,
$(a_0,a_1,\ell_0,\ell_1; c_1,\lambda_0,\lambda_1)$ in total.

The general form of the divergences of the anomalous stress tensor at
the RN event horizon may be found from the power series expansions of
the auxiliary fields there. In the $T^{\theta}_{\ \theta}$ component
there are $s^{-2}$, $s^{-1}$, and $\ln^2 s$ divergences. The leading
$s^{-2}$ behavior in all components of $T^a_{\ b}$ in coordinates
(\ref{static}) may be understood from the approximate conformal
symmetry which applies near the horizon and the conformal weight
of the stress tensor. Namely, since the near horizon geometry is
conformally flat, with $e^{-\varphi_h \pm t/r_+}$ the conformal
transformation to locally flat space, distances scale like
$e^{\frac{\varphi_h}{2}} = f^{\frac{1}{2}}$, while energies scale like
$f^{-\frac{1}{2}}$, and energy densities like $f^{-2} \sim s^{-2}$
near the non-degenerate RN event horizon.

Requiring the coefficients of the leading $s^{-2}$ and subleading
$\ln^2 s$ and $\ln s$ possible divergences in $\Theta$
to be zero gives three conditions on the integration constants of
the auxiliary fields. A possible $s^{-1}$ divergence in $\Theta$ 
turns out to be linearly dependent on the first three, and
is canceled automatically when these three conditions are imposed.
However from (\ref{intp}) with $L=0$ there remains a possible
$s^{-1}$ divergence in $T^{\theta}_{\ \theta}$ unless $C=0$.
This is not automatic and gives a fourth condition. A fifth and 
final condition comes from the necessity of canceling the $\ln s$ 
divergence in the freely falling frame, $T^{uu}$ of (\ref{Tuucondition}). 
A possible $\ln^2 s$ term in $T^{uu}$ drops out after we have satisfied 
the first three conditions, and gives no further condition. Hence we 
end up with five algebraic relations for the seven constants of
integration, {\it viz.},
\begin{subequations}
\begin{eqnarray}
&&\label{gen1} \hskip -2cm\ell_0(b'\ell_0+2b\lambda_0)  =  0  \qquad (s^{-2} \ {\rm in} 
\ T^{\theta}_{\ \theta}) \\
&&\label{gen2} \hskip -2cm\ell_1(b'\ell_1+2b\lambda_1) = 0  \qquad (\ln^2 s \ {\rm in} 
\ T^{\theta}_{\ \theta})\\
&&\label{gen3} \hskip -2cm(b'\ell_1+b\lambda_1)[3-\epsilon (a_1+\ell_1)+
2(3\epsilon-1)\ell_0]+
b\ell_1[3\epsilon - \epsilon (c_1+\lambda_1)+2(3\epsilon -1)\lambda_0] =  0 \quad (\ln s\ 
{\rm in} \ T^{\theta}_{\ \theta}) \\
&&\label{gen4} \hskip -2cm b(6\epsilon \ell_0+ 6 \lambda_0+ 4\epsilon \ell_1\lambda_0
+ 4 \epsilon \ell_0 \lambda_1- \epsilon \lambda_1)
+b'(6\ell_0-\epsilon \ell_1 + 4\epsilon \ell_0\ell_1) = 0
\qquad (C=0)\\
&&\label{gen5} \hskip -2cm b[18\epsilon(1-\epsilon)\ell_0 + 9\epsilon(\epsilon-1)\ell_1
+ 3(1-4\epsilon +3\epsilon^2)\lambda_1 -
(3-12\epsilon+20\epsilon^2)(\ell_1\lambda_0+\ell_0\lambda_1)]\nonumber\\
&&\hskip-1cm + b'[3(1-4\epsilon+3\epsilon^2)\ell_1 - (3-12\epsilon+20\epsilon^2)\ell_1\ell_0]=0
\quad (\ln s \ {\rm in} \ T^{uu})
\end{eqnarray}
\label{RNfinT}
\end{subequations}
\noindent
where $\epsilon$ is given by (\ref{epsdef}). The solutions of these conditions
will be discussed in the next section.

\subsection{ERN Spacetime}

When $Q=M$, $r_+ = r_- = M$, the horizon becomes degenerate, and the
RN metric function $f(r)$ goes to zero quadratically as $r
\rightarrow r_{\pm}$, ($s \rightarrow 0$). This quite different
behavior of the spacetime near the horizon is reflected also in the
behavior of the conformal differential operator $\Delta_4$ and its
solutions. Referring back to (\ref{leadpower}) we observe that when
$\epsilon =0$ the first two terms vanish identically, and the
coefficient of the $s^{\gamma +2}$ term becomes the leading one,
with coefficient $(\gamma+1)(\gamma +2) - 2 = \gamma(\gamma+3)$.
Thus the more singular behavior $\gamma = -3$ is allowed by
(\ref{leadpower}), and the structure of the divergent terms in the
solutions to the auxiliary field equations and stress tensor
(\ref{Tanom}) becomes quite different in the ERN limit. For this
reason the number of conditions necessary for the stress tensor
(\ref{Tanom}) to remain finite increases, and the conditions become
more stringent.

Because of the $s^{-3}$ leading behavior allowed by (\ref{sphphi}) for
$\varphi'$ and $\psi'$, the power series expansions of the auxiliary
fields in the ERN case are of the form,
\begin{subequations}
\begin{eqnarray}
\varphi& = & \sum_{n=-2}a_ns^n + \sum_{n=0}\ell_ns^n \ln s \\
\psi & = & \sum_{n=-2}c_ns^n + \sum_{n=0}\lambda_ns^n \ln s
\end{eqnarray}
\end{subequations}
\noindent instead of (\ref{phiseries}) and (\ref{psiseries}).
Substitution of these series into the differential equations for
$\varphi$ and $\psi$ shows that $(a_{-2},a_{-1},a_0,a_1)$ and
$(c_{-2},c_{-1},c_0,c_1)$ are eight free integration constants, with
$c_0$ again playing no role. The logarithmic terms and all the other
coefficients are related to this set of seven coefficients by recursion
relations. Since the leading behaviors of $\varphi$ and $\psi$ are
now $s^{-2}$ as $s \rightarrow 0$, the stress tensor (\ref{Tanom}) gives
$T^{\theta}_{\ \theta}$ with $s^{-4}$,$ s^{-3}$, $s^{-2}$, $s^{-1}$,
$s^{-1}\ln s$, $\ln^2 s$ and $\ln s$ divergent terms. The leading
$s^{-4}$ behavior again can be understood from the conformal weight
of the stress tensor under conformal transformations near the
horizon, since $f^{-2} \sim s^{-4}$ in the ERN case.

It turns out that the conditions for removing the leading and
all subleading divergent behaviors in the auxiliary field $T^a_{\ b}$ are not
independent, and only {\it four} independent conditions on the integration
constants are sufficient. We can choose these four to be:
\begin{subequations}
\begin{eqnarray}
&&\label{ext1} a_{-2}(2bc_{-2}+b'a_{-2}) =0  \qquad (s^{-4}\ {\rm in}\ \Theta)\\
&&\label{ext2} b(-10a_{-2}c_{-2}+a_{-2}c_{-1}+a_{-1}c_{-2})
+b'(-5a^2_{-2}+a_{-2}a_{-1}) = 0 \qquad  (s^{-3} \ {\rm in} \ \Theta ) \\
&&\label{ext3} b\left[3c_{-1}-2a_{-2}(9+3c_{1}+4c_{-1}-64c_{-2})-
c_{-2}(36+6a_{1}+8a_{-1})\right]+\\ \nonumber
&&\qquad +b'\left[3a_{-1}-a_{-2}(36+8a_{-1}+6a_{1}-64a_{-2})\right] =0
\qquad (s^{-1}\ {\rm in}\ \Theta)\\
&&\label{ext4} a_{-1}(2bc_{-1}+b'a_{-1})=0 \qquad (\ln s \ {\rm in}
\ \Theta)
\end{eqnarray}
\label{ERNfinite}
\end{subequations}
\noindent In particular the last of these conditions removes all logarithmically
divergent terms from $\Theta$.

The requirement (\ref{Tuucondition}) that $T^{uu}$ be finite gives
two additional conditions, {\it viz.},
\begin{subequations}
\begin{eqnarray}
&&\label{ext5}  b(-15+15a_{-1}-96a_{-2}+13c_{-1}-104c_{-2})+\nonumber\\
&&\qquad+b'(13a_{-1}-104a_{-2})=0 \qquad (s^{-1}\ {\rm in}\ T^{uu}) \\
&&\label{ext6} b\left[a_{-2}(486-32c_{-1})+
a_{-1}(3c_{1}-72+8c_{-1}-32c_{-2})+27+3c_{-1}(a_{1}-10)+288c_{-2}\right]\nonumber\\
&&\qquad
+b'\left[4a_{-1}^2+288a_{-2}-a_{-1}(30-3a_{1}+32a_{-2})\right]=0
\qquad (\ln s\ {\rm in}\ T^{uu})
\end{eqnarray}
\label{ERNuufinite}
\end{subequations}

\vskip -1cm \noindent Hence we have six independent conditions on
the seven active integration constants of the auxiliary fields in
order to obtain a fully finite stress tensor from the trace anomaly
in freely falling coordinates in the ERN case. Despite the
discontinuously singular behavior of the geometry as $Q\rightarrow M$,
the more singular behavior of the auxiliary fields in this limit,
and the greater restrictiveness of the finiteness conditions
(\ref{ERNfinite})-(\ref{ERNuufinite}) compared to (\ref{RNfinT}), it
is still possible for the approximation algorithm for the stress
tensor based on the anomalous effective action (\ref{Sanom}) to
yield a fully finite stress tensor on the ERN horizon. This is
qualitatively different from all previous approximation schemes.

\section{Comparison with Previous Approximations and Numerical Results}

Since the method of computing the stress tensor from the anomaly action
with the use of auxiliary fields is relatively new, it is interesting
to compare it to previous approximation methods, as well as direct
numerical evaluations of the renormalized $\langle T^a_{\ b} \rangle$
whenever the latter are available. An analytic approximation to 
$\langle T^a_{\ b} \rangle$ for thermal states of conformal fields 
in non-conformally flat static spacetimes was derived by Page \cite{Page}, 
Brown and Ottewill \cite{BroOtt}, and Brown, Ottewill and Page \cite{BOP}. 
This approximation is based upon the semi-classical approximation to the 
proper time heat kernel \cite{BekPar}, and the properties of the exact 
one-loop effective action under the conformal transformation,
\begin{equation}
g_{ab} = e^{2\omega}\, \tilde g_{ab}\,.
\label{Weyl}
\end{equation}
For a classical conformal field the dependence of the exact effective
action upon $\omega$ is determined completely by the trace anomaly
in the form,
\begin{equation}
S_{exact}[g] = S_{exact}[\tilde g] + b A[\omega;g] + b' B[\omega; g]\,,
\label{conAB}
\end{equation}
where $A[\omega;g]$ and $B[\omega;g]$ are known (non-linear)
functionals of $\omega$ and $g_{ab}$ which are given in
\cite{BroOtt,BOP}. If the conformal transformation $e^{2\omega}$ is
chosen to be $f(r)$, for the static, spherically symmetric line
element (\ref{static}), then the conformally transformed metric
$\tilde g_{ab}$ becomes the ultrastatic, optical metric, for which
$\tilde g_{tt} = -1$. If, in addition, the original physical metric
$g_{ab}$ is that of a static Einstein space ({\it i.e.} one for
which $R_{ab} = \Lambda g_{ab}$), then the invariants appearing in
the anomaly,
\begin{subequations}
\begin{eqnarray}
&& \frac{2}{\sqrt {-g}}\, g_{ab}\, \frac{\delta A}{\delta g_{ab}} =
\left[F + \frac{2}{3} \sq R \right]_{g= \tilde g} = 0 \,,\\
&& \frac{2}{\sqrt {-g}}\, g_{ab}\,\frac{\delta B}{\delta g_{ab}} =
E \big\vert_{g= \tilde g} = 0\,,
\end{eqnarray}
\label{PBOAB}
\end{subequations}

\vskip -1cm \noindent vanish for the conformally related ultrastatic
metric $\tilde g_{ab}$, and the $A$ and $B$ terms in (\ref{conAB})
reproduce the correct trace for the physical metric $g_{ab}$. Thus, with
this choice of $\omega$ it is consistent simply to neglect
$S_{exact}[\tilde g]$ in (\ref{conAB}), and approximate the full
$\langle T^{a}_{\ b} \rangle$ by the terms coming from $A[\omega;g]$
and $B[\omega;g]$ which are known analytically.

Because it is also based on the form of the trace anomaly, the Page-Brown-Ottewill
(PBO) approximation is related to the approximation scheme based on the
auxiliary field effective action of the anomaly. Indeed, by making use of
the conformal transformation property,
\begin{equation}
\sqrt{-\tilde g} \tilde R^2 = \sqrt{-g}\left[ R + 6(\sq \omega - g^{ab}
\partial_a\omega\partial_b\omega)\right]^2
\end{equation}
it is not difficult to show that
\begin{equation}
b A[\omega;g] + b' B[\omega; g] = - \Gamma_{WZ}[g; -\omega] + \frac{(b+b')}{18}
\int d^4x (\sqrt{-\tilde g} \tilde R^2 - \sqrt{-g} R^2)\,,
\label{Rsq}
\end{equation}
where
\begin{equation}
\Gamma_{WZ}[g; -\omega] = S_{anom}[g] - S_{anom}[\tilde g = e^{-2\omega} g]
\end{equation}
is the Wess-Zumino effective action for the anomaly obtained in
\cite{MazMot,MotVau}. Since $\Gamma_{WZ}$ is a quadratic functional
of $\omega$, the relation (\ref{Rsq}) shows that the complicated
cubic and quartic terms of the PBO effective action are simply the
result of adding an $R^2$ term to $\Gamma_{WZ}$, with a
corresponding $b'' \sq R$ term in the trace anomaly (\ref{tranom}).
The PBO effective action $W_{PBO}[g]$ is related then to the anomaly
effective action $S_{anom}$ of (\ref{Sanom}) by
\begin{eqnarray}
&&W_{PBO}[g] = W_{PBO}[\tilde g] +  \frac{(b+b')}{18}
\int d^4x \sqrt{-\tilde g} \tilde R^2 - S_{anom}[\tilde g] \nonumber\\
&& \qquad + S_{anom}[g] -  \frac{(b+b')}{18} \int d^4x \sqrt{-g} R^2\,.
\end{eqnarray}
This consistent with the decomposition (\ref{Sexact}), in which
all terms that depend only upon $\tilde g_{ab}$ are viewed as
invariant under the local Weyl transformation (\ref{Weyl}). Because
of the different linear combination of invariants in (\ref{PBOAB})
from the $b$ and $b'$ terms in (\ref{tranom}), the PBO conformal
transformation is given by an $\omega$ which satisfies
\begin{equation}
4 \Delta_4 \omega = E - F - \frac{2}{3} \sq R\,,
\end{equation}
and therefore corresponds to a particular linear combination ({\it i.e.}
$\varphi - \psi$) of the auxiliary fields, with a particular choice
of homogeneous solution.

Although the PBO approximation is related to the one based on
$S_{anom}$, there are two important differences. First, in the PBO
approximation the conformal transformation $e^{2\omega}$ to the
ultrastatic metric is {\it fixed} up to a linearly time dependent
term in $\omega$, by the requirement that the invariants in
(\ref{PBOAB}) vanish in the conformally transformed spacetime. Hence
there is only one parameter, namely the coefficient of linear time
dependence free in $\omega$, which is far fewer than the seven
integration constants corresponding to the freedom to adjust the
state dependent and Weyl invariant part of the effective action in
the auxiliary field method. Secondly, and even more importantly, the
invariants in (\ref{PBOAB}) vanish in the conformally transformed
spacetime only in rather special cases, such as if the original
metric $g_{ab}$ is that of a static Einstein space. Although Zannias
suggested a modification of the PBO ansatz to account for the
correct trace in non Einstein spaces \cite{Zan}, the non-vanishing
of the trace in the conformally related space with metric $\tilde
g_{ab}$ means that the original PBO rationale for ignoring
$S_{exact}[\tilde g]$ in (\ref{conAB}) no longer applies. Hence
there is no especially good reason why this modification of the PBO
approximation should be accurate, or even finite on the event
horizon. If one tries instead to find a spacetime where the
conditions (\ref{PBOAB}) for the vanishing of the trace are
satisfied, then one is faced with solving {\it two} non-linear 
conditions for a single $\omega$. Hence one would not expect that a
simultaneous solution of both conditions (\ref{PBOAB}) exists at all
for $\tilde g_{ab} = e^{-2\omega} g_{ab}$, for general $g_{ab}$.

In the auxiliary field method, these difficulties are removed
completely. The {\it two} independent $\varphi$ and $\psi$ fields
satisfy the {\it linear} equations (\ref{EFtraces}), for which
solutions are guaranteed to exist and can be found in {\it any}
spacetime. Although $e^{-\varphi}$ may be regarded as a conformal
transformation to a spacetime where $E - \frac{2}{3} \sq R$
vanishes, there is no such conformal interpretation for the second
auxiliary field $\psi$, and none is required. The auxiliary scalars
and their stress tensor simply encode the same information about the
full non-local trace anomaly in a local, generally covariant form,
and there is no conformally related $\tilde g_{ab}$ enjoying a
privileged status over any other. This is clear also from the
freedom to add arbitrary homogeneous solutions to (\ref{EFtraces}),
corresponding to different choices for the conformal image $\tilde
g_{ab}$, and different Weyl invariant parts of the effective action.
This additional freedom in the linear system (\ref{EFtraces}) makes
an exploration of a much wider class of states possible with the
auxiliary field algorithm, with different state-dependent traceless
contributions to the stress tensor, all consistent with the correct
trace anomaly. It is of course the much larger parameter space
available in the auxiliary field method which makes it possible to
find regular stress tensors on both the RN and ERN event horizons.

The PBO approximation was later rederived in a different way by
Frolov and Zel'nikov (FZ), by carrying out an analysis of possible
terms in the effective action in general spacetimes with a static
Killing vector field, $K_a = \frac{\partial}{\partial t}$
\cite{FroZel}. Although rather different in methodology, the FZ
approach also applies only to static spacetimes and makes use of the
conformal transformation properties (conformal weights) of the
various terms in the stress tensor. As a result, although the FZ
approach is not limited {\it a priori} to Einstein spaces, the
effective action they obtain is in fact equivalent to that of PBO
(as modified by Zannias), up to local $F=C_{abcd}C^{abcd}$ and the
$(\sqrt{-\tilde g} \tilde R^2 - \sqrt{-g} R^2)$ terms appearing in
(\ref{Rsq}), which are allowed to have arbitrary coefficients
unrelated to the anomaly. These particular Weyl invariant terms are
mildly behaved on the event horizon, and hence cannot be used to
cancel any divergences present in the PBO approach. Hence when applied 
to the non-Ricci flat Reissner-Nordstr\"{o}m geometry, the FZ 
approximation suffers from the same limitation as that of PBO, namely, 
both predict a logarithmically divergent stress tensor on the RN event
horizon, and both a linear and a logarithmic divergence on the ERN
horizon.

From our present vantage point, the interesting feature of the FZ
approach is the central role of the static Killing field $K_a$. The
FZ approach underlines the fact that $\langle T^a_{\ b} \rangle$
generally depends upon {\it global} invariant functions of $K_a$ as
in (\ref{Kill}), in addition to strictly local invariants such as
$F$, $E$ and $R^2$. Since global invariants such as (\ref{Kill}) may
diverge on the event horizon, and $\langle T^a_{\ b} \rangle$
generally is a function of these invariants, explicitly so
in the FZ approach, it is clear that requiring such divergences 
to be absent in $\langle T^a_{\ b} \rangle$ is a dynamical 
restriction on the quantum state, not at all required by general 
coordinate invariance.

It was shown by Howard and Candelas~\cite{HowCan,How} for the
conformally invariant spin $0$ field in
Schwarzschild spacetime that the WKB approximation can be used to
write the stress-energy tensor in terms of the PBO approximation for
each field plus a term containing mode sums that must be computed
numerically.  A similar result was obtained for the massless spin $1$ field
by Jensen and Ottewill~\cite{JenOtt}.  Anderson, Hiscock, and Samuel~\cite{AnHiSa} 
(AHS) showed that if the WKB approximation in the high
frequency limit is used for the radial modes of the Euclidean Green's
function for a massless scalar field with arbitrary curvature coupling
in a general static spherically symmetric spacetime, then a conserved
stress-energy tensor results which in the case of conformal coupling has
a trace equal to the trace anomaly. For the case of conformal coupling this
stress-energy tensor is equivalent to that derived by FZ if the
three arbitrary constants in their derivation are set to zero.  A
similar approximation was derived by Groves, Anderson, and
Carlson~\cite{GrAnCa} (GAC) for the massless spin $1/2$ field.  In 
this case the approximation is equivalent to that of FZ if their
arbitrary constants have the values $q_1^{(0)} = q_2^{(0)} = 0$ and
$q_1^{(2)} = 1/144$.

Huang's evaluation of the stress tensor for a conformally invariant
scalar field in the PBO/FZ approximation \cite{Hua}, apparently correcting
an error in \cite{Zan} for the ERN case, shows this same logarithmic
divergence of (\ref{Tuu}) for $Q<M$, which becomes a combination of a
linear and logarithmic divergence in the ERN case.\footnote{The third
footnote of Ref. \cite{Hua} gives an incorrect criterion for
finiteness of the stress tensor on the horizon in freely falling
coordinates, replacing one factor of $f(r)$ in (\ref{Tuu}) by
$f^{\frac{1}{2}}(r)$. For this reason the conclusions drawn in
\cite{Hua} are not warranted by the evaluation of $T^a_{\ b}$
given.}  Not surprisingly the same divergences were found for the
AHS and GAC approximations~\cite{AnHiSa,CHOAG}. Thus all pre-existing
analytic approximations to $\langle T^a_{\ b}\rangle$ diverge
at least logarithmically in the general RN spacetime, and linearly
in the ERN case.

\subsection{Schwarzschild Spacetime}

The six conditions (\ref{ScondA}) and (\ref{ScondB}) guaranteeing
finiteness of the stress tensor on the Schwarzschild horizon can be
satisfied in several different ways.
The simplest possibility is \cite{MotVau}
\begin{subequations}
\begin{eqnarray}
&& (2b + b') c_{_H}^2 + \eta (2b \eta' + b'\eta) = 0\,,\\
&& bd_{_H} = -(b + b') c_{_H}\,,\\
&& q = q' = 2 \,,\\
&& b'\eta q + b(q\eta' + q'\eta) = 0\,,
\end{eqnarray}
\label{fincond}
\end{subequations}

\vskip-1cm
\noindent
which requires fixing only five parameters to satisfy all six
conditions (\ref{ScondA}) and (\ref{ScondB}). This is
because the choice $q=q'=2$ eliminates all the logarithmic terms
in the auxiliary fields at the horizon (hence all $\ln^2 s$ and $\ln s$
terms in $T^a_{\ b}$) and simplifies the remaining conditions
considerably. The three parameter subset of the original eight
parameter family of spherically symmetric auxiliary fields is
certainly not generic, but still leaves considerable freedom to fit
the finite values of the stress tensor at $r=2M$ and/or
$r=\infty$ in the regular Hartle-Hawking-Israel state~\cite{HarHaw}. This
is the choice which was studied in some detail in \cite{MotVau}. The other
possible ways of solving the finiteness conditions are listed
in the Appendix. Approximate stress tensors for the Unruh~\cite{Unr} state may be
obtained by replacing the $L=0$ condition (\ref{lum}) by its correctly
normalized value.

The PBO/FZ approximation works quite well for the conformally invariant scalar 
field in the Hartle-Hawking-Israel state in Schwarzschild spacetime, although 
the method based on the auxiliary field effective action of the anomaly accounts 
for the behavior of the stress tensor more accurately in states with divergent 
stress tensors on the horizon, such as the Boulware state in Schwarzschild
spacetime.  For the spin $1/2$ and spin $1$ fields the PBO/FZ approximation 
works less well~\cite{CHOAG,JenOtt}.

Balbinot, Fabbri, and Shapiro~\cite{BFS} (BFS) were the first to use
two auxiliary fields to study the properties of the approximate
stress-energy tensor based on the anomaly in Schwarzschild spacetime. 
Thus their approach comes the closest of any previous one to the present 
work. Their two auxiliary fields $(\phi, \psi)$ are related to ours by 
the linear combinations, $-\sqrt{-b'}\varphi + b\psi/\sqrt{-b'}$ and
$b/\sqrt{-b'} \psi$ respectively. However, BFS applied conditions to
each of these two linear combinations of auxiliary fields
separately, rather than searching for the more general solutions of
the finiteness conditions (\ref{ScondA}) and (\ref{ScondB}) on the
Schwarzschild horizon.  This amounts to fixing the traceless terms
in the stress tensor with fewer free parameters than in the present
approach. For this reason they were not able to reproduce some
features of the exact stress tensor in the Hartle-Hawking-Israel state,
such as the correct value of $\langle T^a_{\ b} \rangle$ on the
Schwarzschild event horizon or at infinity. In the case of the
present formulation, after all four finiteness conditions
(\ref{ScondA})-(\ref{ScondB}) are satisfied, there remains a three
parameter family of finite stress tensors. Hence both the values on
the horizon and at infinity can be adjusted to their correct values
by a suitable choice of the integration constants.

As pointed out by BFS, the subdominant terms in the stress tensor at
infinity are also not reproduced by the stress tensor of the
anomaly, and this undesirable feature persists in our approach.
Clearly this is because the anomaly action is not equal to the full
quantum effective action, differing from it by Weyl invariant terms,
as in (\ref{Sexact}). In regular states these give rise to
additional traceless terms in the stress tensor, which would be
expected to be of the same order as those in (\ref{Tanom}). One
possibility for improvement is to add the Weyl invariant term,
\begin{equation}
S_{inv} [g; \chi] = k\, \int\,\sqrt{-g}\,d^4x\, \left(-\frac{1}{2} \chi
\Delta_4 \chi + F\chi \right)
\label{Sinv}
\end{equation}
to the total effective action. If we added this term to $S_{anom}$,
we would have a third auxiliary field, denoted here by $\chi$, with
four more integration constants to serve as free parameters in the
stress tensors of the RN and ERN spacetimes. Although this may allow
for more accurate fitting of the numerical results for
$\langle T^a_{\ b} \rangle$ in the RN and ERN cases, we do not
pursue this possibility here, and thus set $k=0$ in our approximation.

\subsection{General RN Spacetime}

When the PBO/FZ, AHS, and GAC approximations are applied to the
$Q>0$ RN spacetimes, they predict that $\langle T^a_{\ b} \rangle$
diverges logarithmically on the RN event horizon for $Q<M$ and
linearly and logarithmically for $Q=M$. The reason for the
divergence of the PBO/FZ approximation is that the semi-classical
approximation for the heat kernel of \cite{BekPar} fails as
$r\rightarrow r_+$. Thus for massless fields all previous approximations 
are both highly specialized to certain specific classes of spacetimes, 
and also lead inevitably to divergences on the event horizons of
non-Einstein static spaces, such as the Reissner-Nordstr\"{o}m
metric. The present approach based on the auxiliary field form of
the effective action of the anomaly does not rely on a
WKB approximation, and hence does not suffer from
these limitations.

The direct method of evaluating $\langle T^a_{\ b} \rangle$ from the
wave equations of the underlying conformal field theory has been
carried out for both scalar and spinor fields in RN spacetimes with
$0 < Q \le M$~\cite{AnHiSa,AHL,CHOAG}. These direct evaluations show no
evidence of a divergence as the horizon is approached.  In the
following plots we show the results of these exact numerical evaluations
of $\langle T^a_{\ b} \rangle$ and the approximation of the present paper
for an RN spacetime with $|Q|/M = 0.99$.

The approximation based on the auxiliary field effective action is
qualitatively better than that of PBO/FZ in all $Q>0$ RN regular
states in that it allows for a finite stress tensor on the event
horizon, showing roughly comparable features to those already
encountered in the Schwarzschild case in \cite{MotVau}. The main
difference from the $Q=0$ Schwarzschild case to the $0<Q \le M$ RN
cases is the loss of the integration constants $\eta$, $\eta'$ in
the latter cases, since such a linear time dependence in $\varphi$
or $\psi$ leads to time dependence in the stress tensor
(\ref{Tanom}) in the non-Ricci flat RN geometries.

In the present approximation algorithm there are various possible
ways of solving the five finiteness conditions (\ref{RNfinT}). For
example, the first condition (\ref{gen1}) is solved by either
\begin{subequations}
\begin{eqnarray}
&& \ell_0 = 0 \qquad {\rm or} \\
&& \ell_0 = -\frac{2b}{b'} \lambda_0\,.
\end{eqnarray}
\end{subequations}
\noindent
Likewise, independently of this choice, the second condition (\ref{gen2})
is solved by either
\begin{subequations}
\begin{eqnarray}
&& \ell_1 = 0 \qquad {\rm or} \\
&& \ell_1 = -\frac{2b}{b'} \lambda_1\,.
\end{eqnarray}
\end{subequations}
\noindent
When the the remaining three conditions are considered, it becomes apparent
that either $\lambda_0$ and $\lambda_1$ are both zero, or they are both
non-zero. The first option, {\it viz.}
\begin{equation}
\ell_0 = \lambda_0 = \ell_1 = \lambda_1 = 0\,, \qquad {\rm (minimal)}
\label{RNmin}
\end{equation}
we term ``minimal" because it requires only four integration constants
be set to zero in order to satisfy all five finiteness conditions
(\ref{RNfinT}). Since this leaves the most integration constants still
free to adjust, we consider this minimal solution of the finiteness
conditions in order to compare the auxiliary field algorithm to the
direct evaluation of $\langle T^a_{\ b} \rangle$. The other possibilities
for solving the conditions (\ref{RNfinT}) are catalogued in the Appendix.

\begin{figure}
\vskip -0.2in \hskip -0.4in
\includegraphics[angle=90,width=3.4in,clip]
{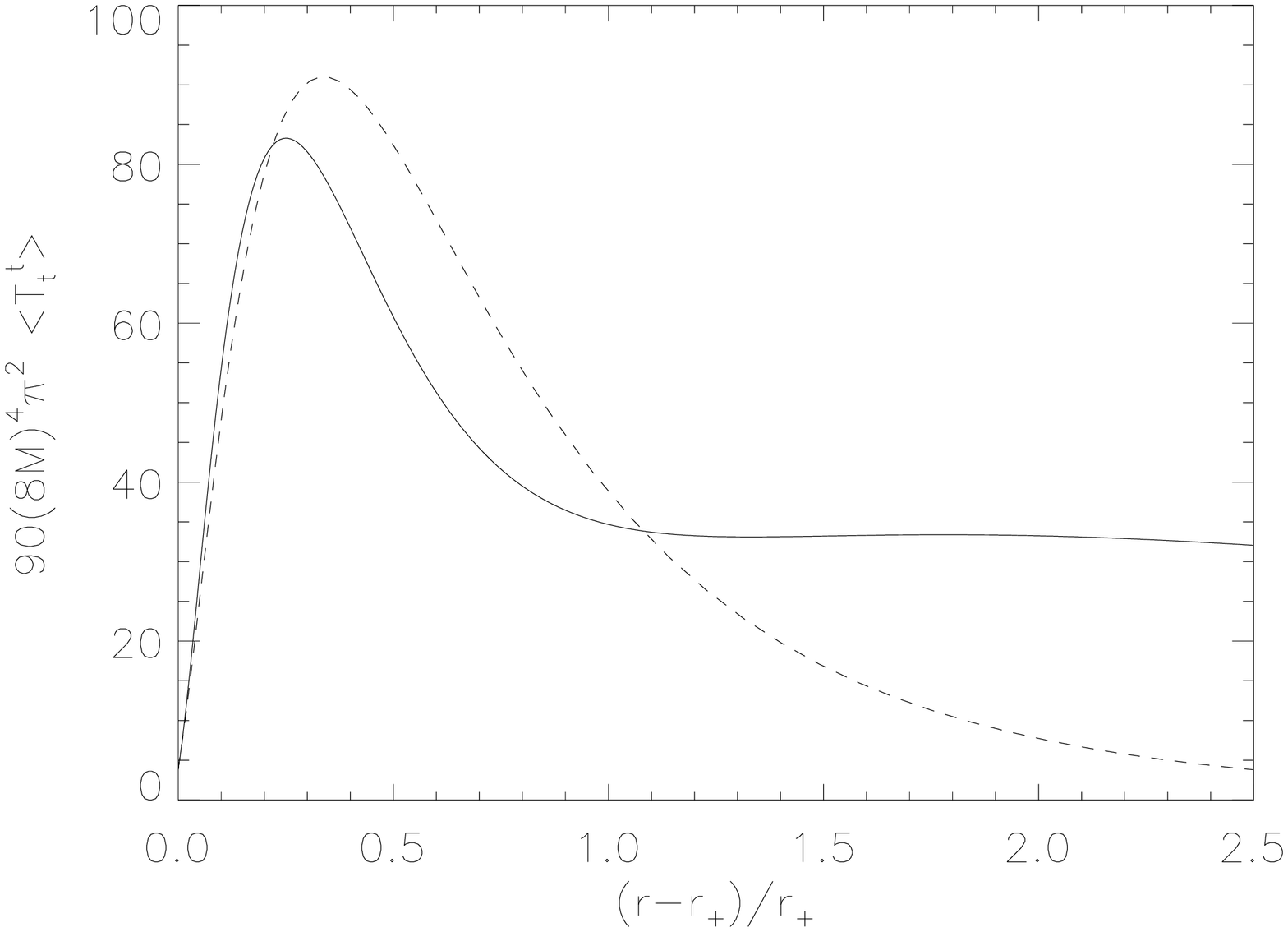}
\vskip -.2in \caption{ The expectation value
$\langle T^t_{\ t}\rangle$ for a conformally invariant scalar field
in the Reissner-Nordstr\"{o}m geometry with $Q=0.99 M$.  The solid
line corresponds to the auxiliary field stress tensor and the dashed
line to the numerically computed exact one.} \label{fig:RNtt0}
\end{figure}

\begin{figure}
\vskip -0.2in \hskip -0.4in
\includegraphics[angle=90,width=3.4in,clip]
{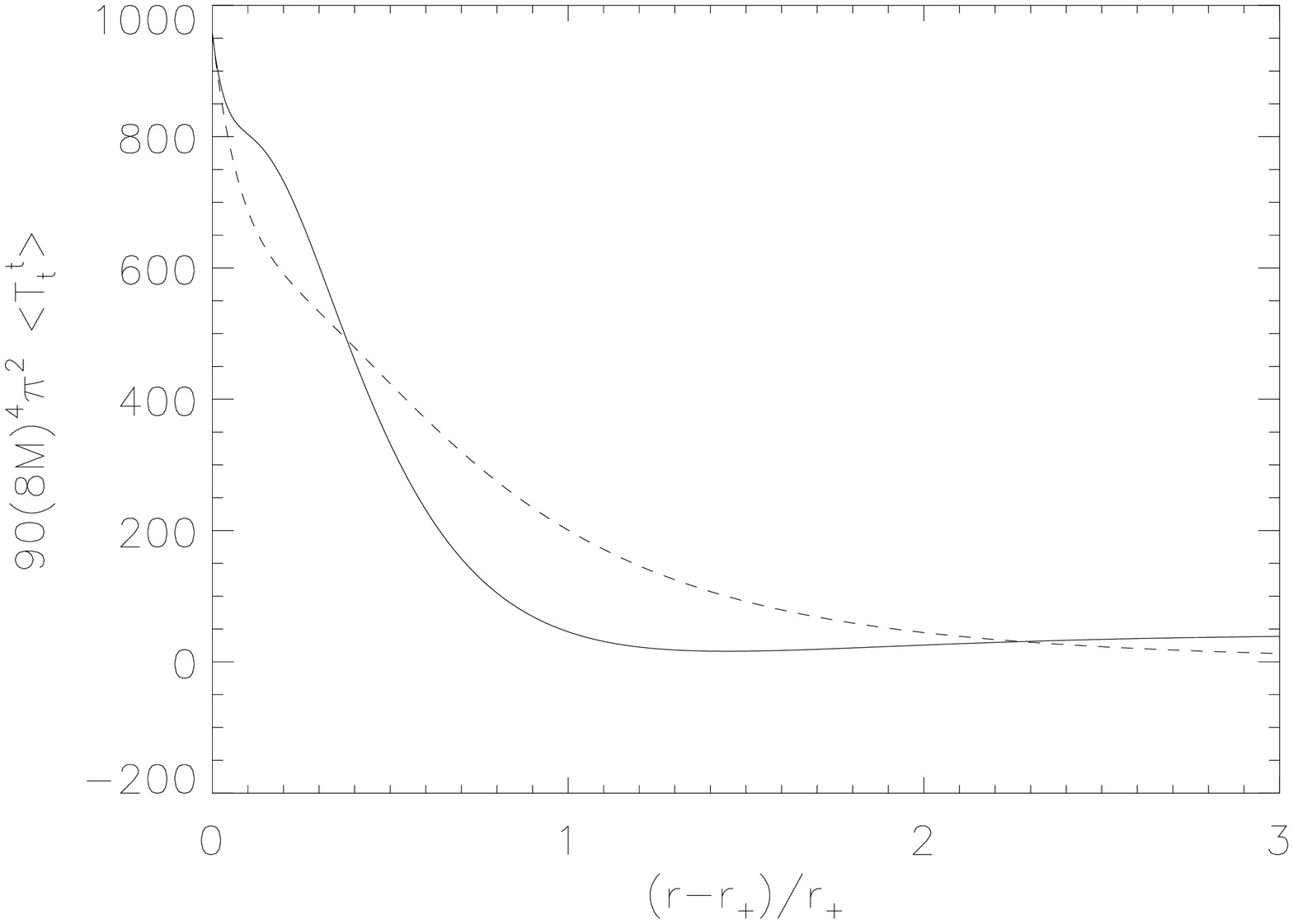}
\vskip -.2in \caption{The expectation value
$\langle T^t_{\ t}\rangle$ for a massless spin $1/2$ field in the
Resisner-Nordstr\"{o}m geometry with $Q=0.99 M$. The solid line
corresponds to the auxiliary field stress tensor and the dashed line
to the numerically computed exact one.} \label{fig:RNtt12}
\end{figure}

\begin{figure}
\vskip -0.2in \hskip -0.4in
\includegraphics[angle=90,width=3.4in,clip]
{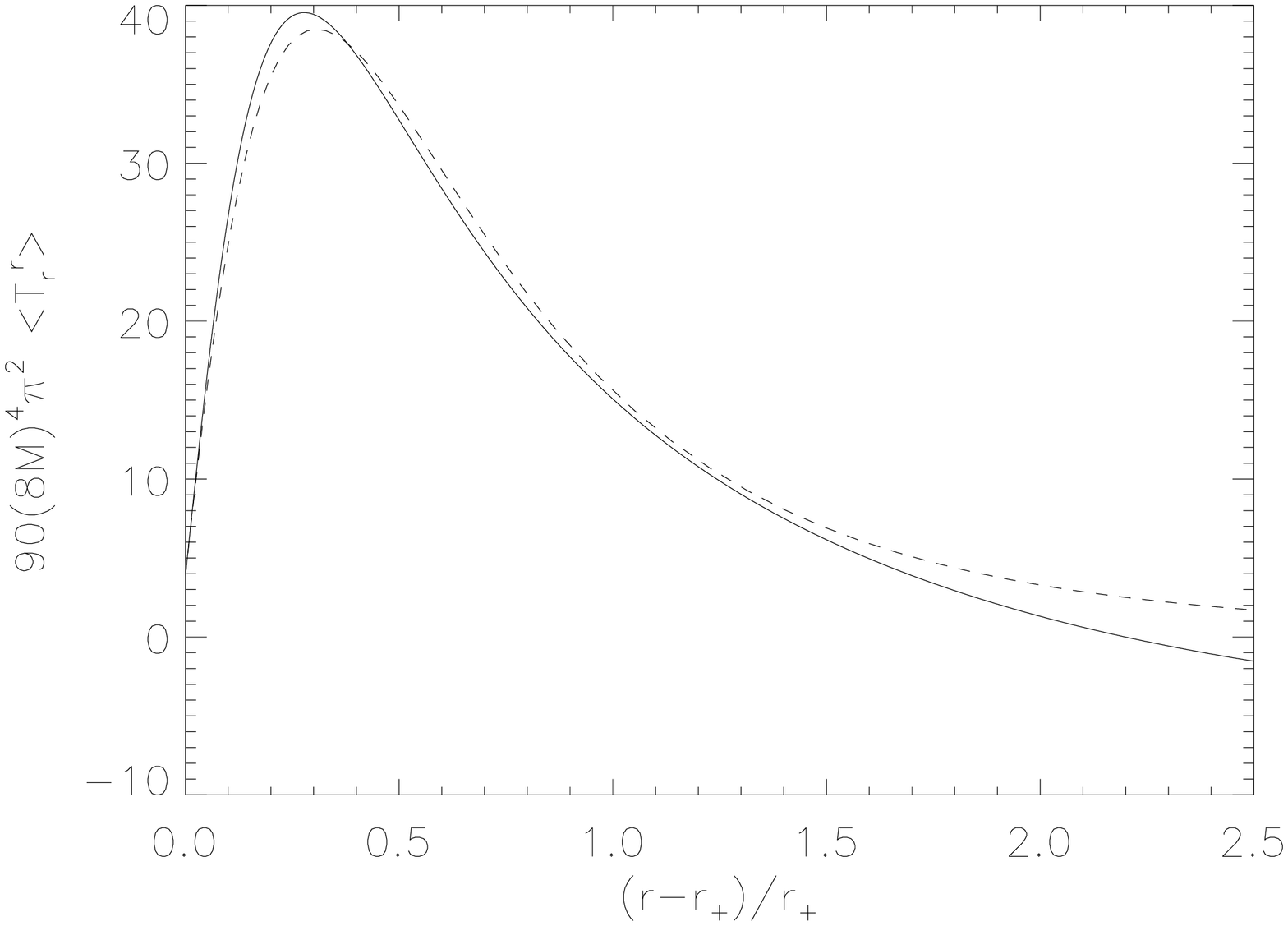}
 \vskip -.2in \caption{The expectation value
$\langle T^r_{\ r}\rangle$ for a conformally invariant scalar field
in the Reissner-Nordstr\"{o}m geometry with $Q=0.99 M$.  The solid
line corresponds to the auxiliary field stress tensor and the dashed
line to the numerically computed exact one.} \label{fig:RNrr0}
\end{figure}

\begin{figure}
\vskip -0.2in \hskip -0.4in
\includegraphics[angle=90,width=3.4in,clip]
{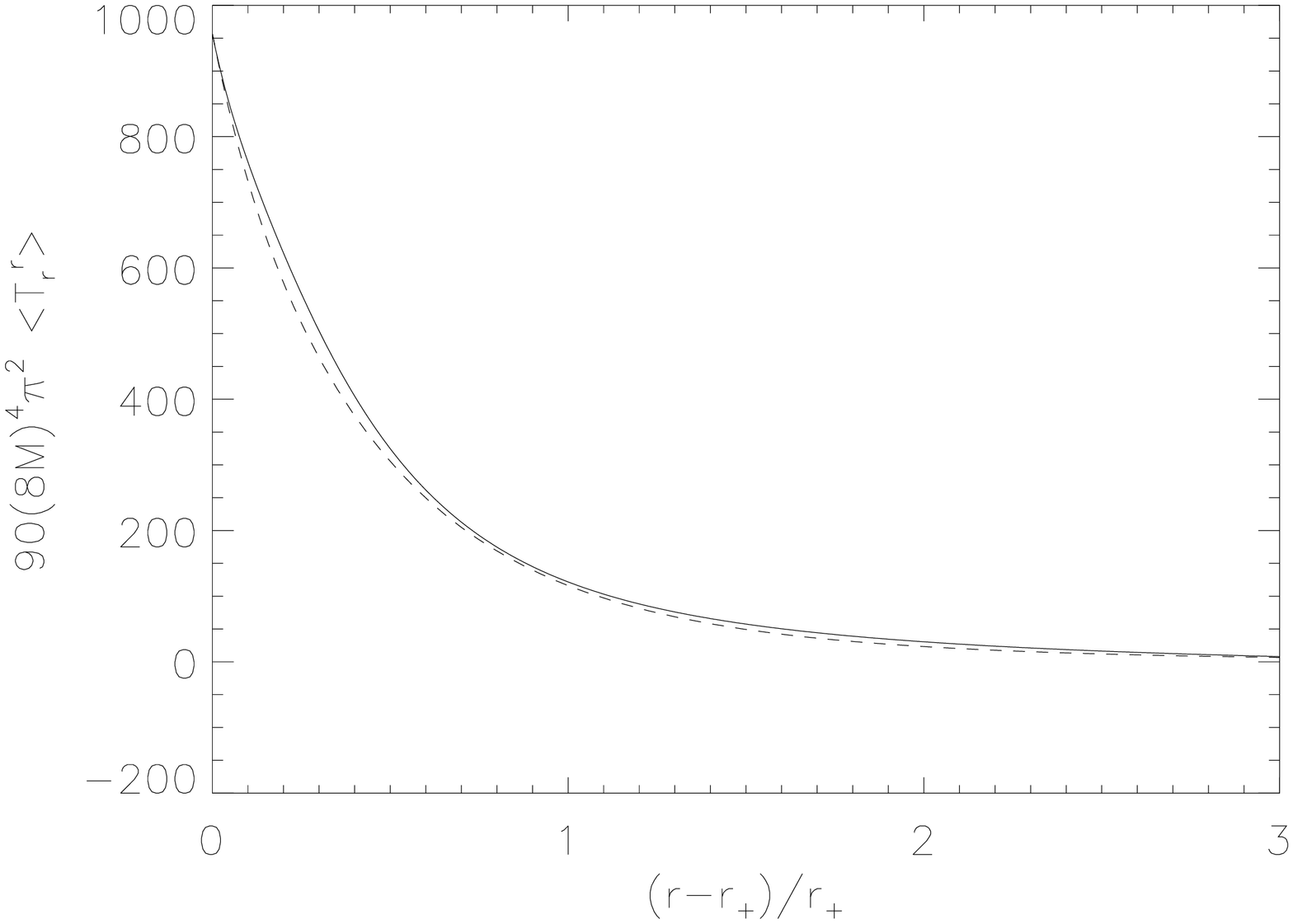}
\vskip -.2in \caption{The expectation value
$\langle T^r_{\ r}\rangle$ for a massless spin $1/2$ field in the
Resisner-Nordstr\"{o}m geometry with $Q=0.99 M$. The solid line
corresponds to the auxiliary field stress tensor and the dashed line
to the numerically computed exact one.} \label{fig:RNrr12}
\end{figure}

The remaining three nontrivial integration constants are $a_0$,
$a_1$, and $c_1$. If the stress tensor is finite on the horizon
then $T^r_{\ r} = T^t_{\ t}$ there.  So once the value of ${T_t}^t$ is
known, the value of $T^{\theta}_{\ \theta}$ on the horizon is fixed by
the value of the trace anomaly there. One can compute the
components of the approximate stress tensor on the horizon from the
general expressions (\ref{Eab}), (\ref{Fab}), and (\ref{Tanom}) by
using the metric (\ref{static}) and metric functions (\ref{fhRN})
along with the expansions (\ref{phiseries}) and (\ref{psiseries}).
One finds that the value of $T^t_{\ t}$ on the horizon depends on
$a_1$ and $c_1$ and the value of $T^{uu}$ on the horizon depends on
$a_0$, $a_1$ and $c_1$.

A numerical code was developed which for a given value of $a_1$
chose the values of $a_0$ and $c_1$ so that ${T_t}^t$ and $T^{uu}$
matched the values previously obtained~\cite{AnHiSa,CHOAG} from exact
numerical computations of the stress tensor on the horizon for the
spin $0$ and spin $1/2$ fields. This code then solves the equations
for the auxiliary fields $\phi$ and $\psi$ and computes the analytic
approximation (\ref{Tanom}) for the stress tensor for various values
of the radial coordinate $r$. Different values of $a_1$ lead to
different behaviors at large values of $r$. The goal is to find the
value of $a_1$ which gives the same energy density at large $r$ as
the field has if it is in the Hartle-Hawking-Israel state.  This is the
best that one can do to find an approximation for the stress-energy
which is finite on the horizon and has the correct energy density at
large values of $r$. One could of course take the opposite approach
and fix the behavior at large $r$ more accurately, but then the
stress-energy would not be regular on the event horizon.

Our results for the spin $0$ field are shown in Figures 1 and 3.  It
can be seen that the approximation is a reasonably good
approximation close to the horizon and at intermediate values of
$r$.  However, at larger values of $r$ the components of the
approximate stress tensor do not approach their asymptotic values
nearly as quickly as they should.  For the spin $1/2$ field there
were actually two values of $a_1$ which gave the correct energy
density at large $r$. However, only one is a good approximation to
the stress-energy tensor near the horizon and that one is shown in Figs. 
2 and 4.

\subsection{ERN Spacetime}

Finally in the ERN case of $Q=M$ the tensor (\ref{Tanom}) is still
able to return a completely finite approximation to $\langle T_a^{\
b} \rangle$ at $s=0$, provided that all six conditions in Eqs.\
(\ref{ERNfinite}) and (\ref{ERNuufinite}) are fulfilled. There are
three algebraically distinct ways of solving these equations, which
are described in detail in the Appendix. The simplest possibility
which is also minimal in the sense that it requires only five
integration constants be fixed in order to satisfy all six
conditions (\ref{ERNfinite}) is the choice,
\begin{subequations}
\begin{eqnarray}
a_{-2} & = & 0 \\
a_{-1} & = & 0\\
a_{1} &=&-\frac{9}{5}\\
c_{-2} &=& \frac{75}{26}\\
c_{-1} &=& \frac{315}{13}
\end{eqnarray}
\label{ERNsoln}
\end{subequations}
\begin{figure}
\vskip -0.2in \hskip -0.4in
\includegraphics[angle=90,width=3.4in]
{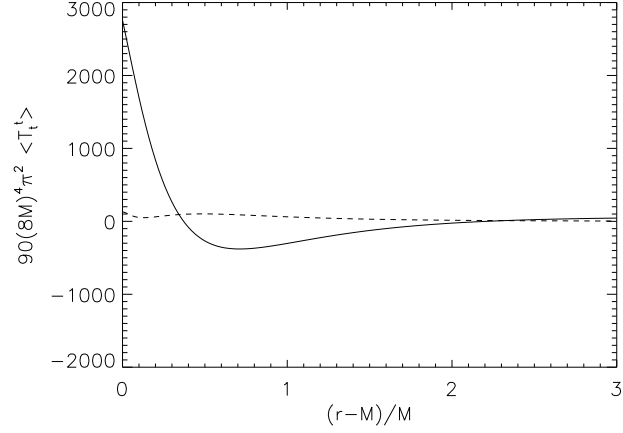}
\vskip -.2in \caption{The
expectation value $\langle T^t_{\ t}\rangle$ for a conformally
invariant scalar field in the extreme Reissner-Nordstr\"{o}m
geometry. The solid line corresponds to the auxiliary field stress
tensor and the dashed line to the numerically computed exact one. }
\label{fig:ERNtt}
\end{figure}

\begin{figure}
\vskip -0.2in \hskip -0.4in
\includegraphics[angle=90,width=3.4in]
{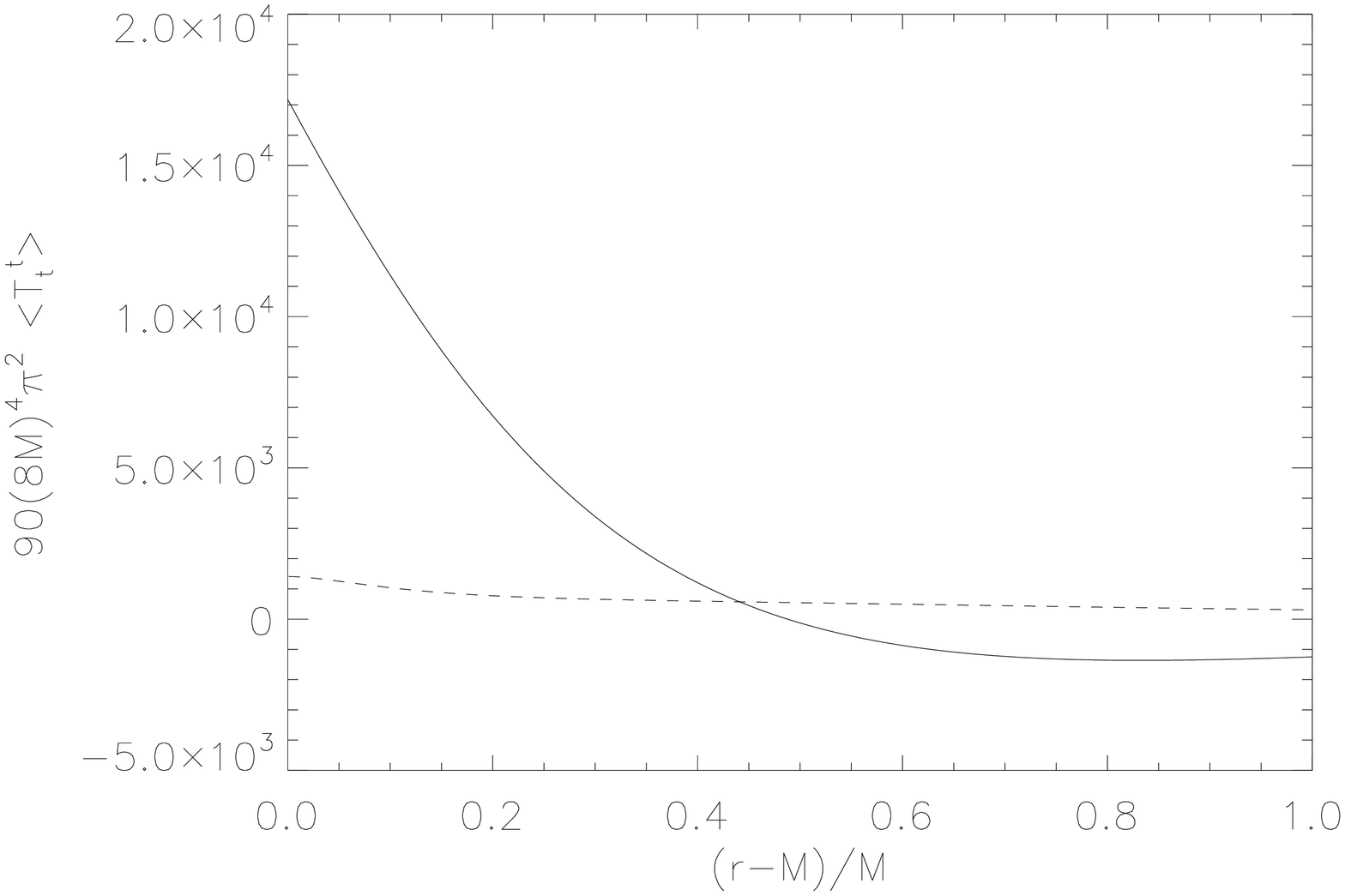}
 \vskip -.2in \caption{The
expectation value $\langle T^t_{\ t}\rangle$ for a massless spin
$1/2$ field in the extreme Reissner-Nordstr\"{o}m geometry. The solid
line corresponds to the auxiliary field stress tensor and the dashed
line to the numerically computed exact one.} \label{fig:ERNtt12}
\end{figure}

\begin{figure}
\vskip -0.2in \hskip -0.4in
\includegraphics[angle=90,width=3.4in]
{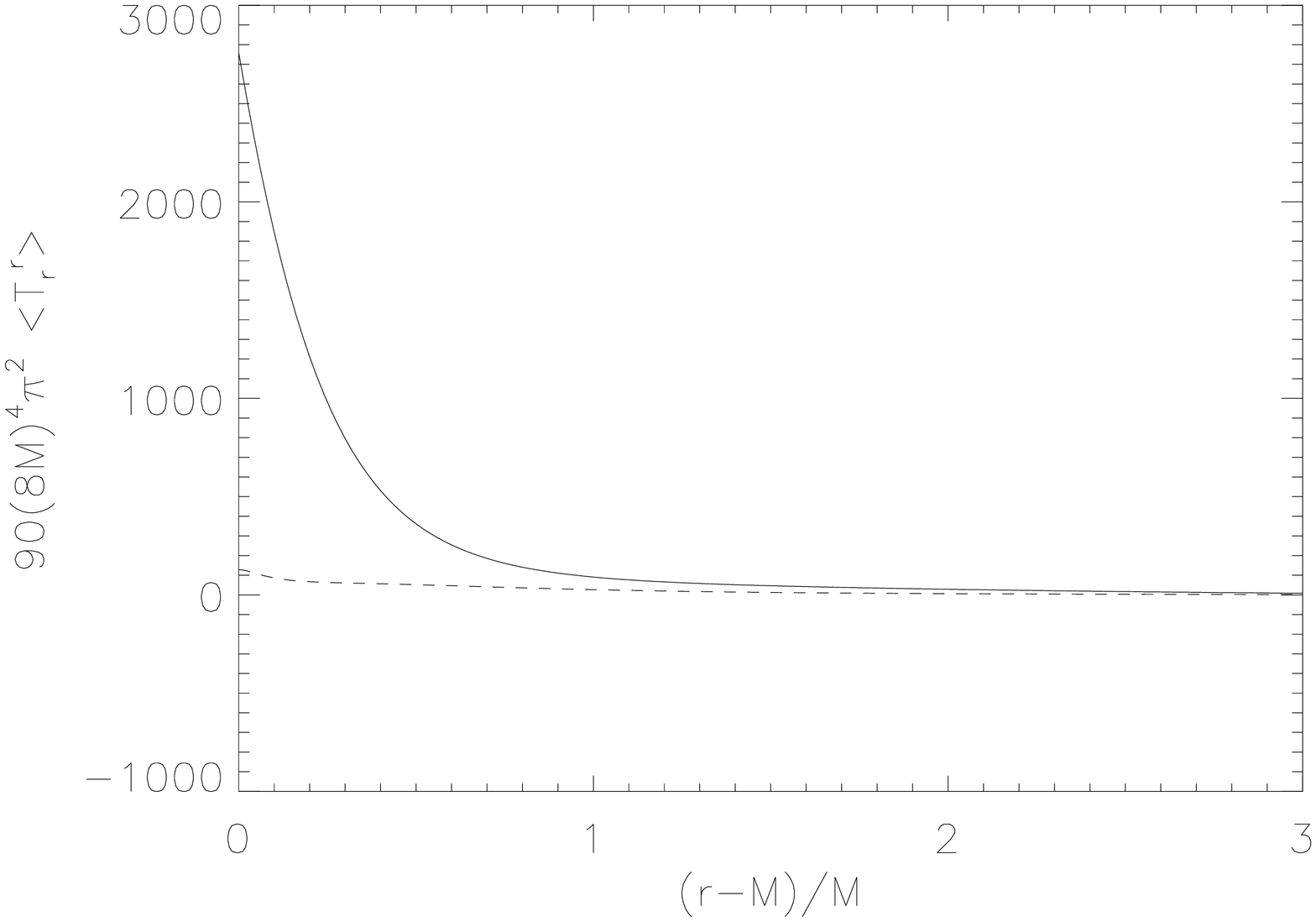}
 \vskip -.2in \caption{The
expectation value $\langle T^r_{\ r}\rangle$ for a conformally
invariant scalar field in the extreme Reissner-Nordstr\"{o}m
geometry. The solid line corresponds to the auxiliary field stress
tensor and the dashed line to the numerically computed exact one. }
\label{fig:ERNTrr}
\end{figure}

\begin{figure}
\vskip -0.2in \hskip -0.4in
\includegraphics[angle=90,width=3.4in]
{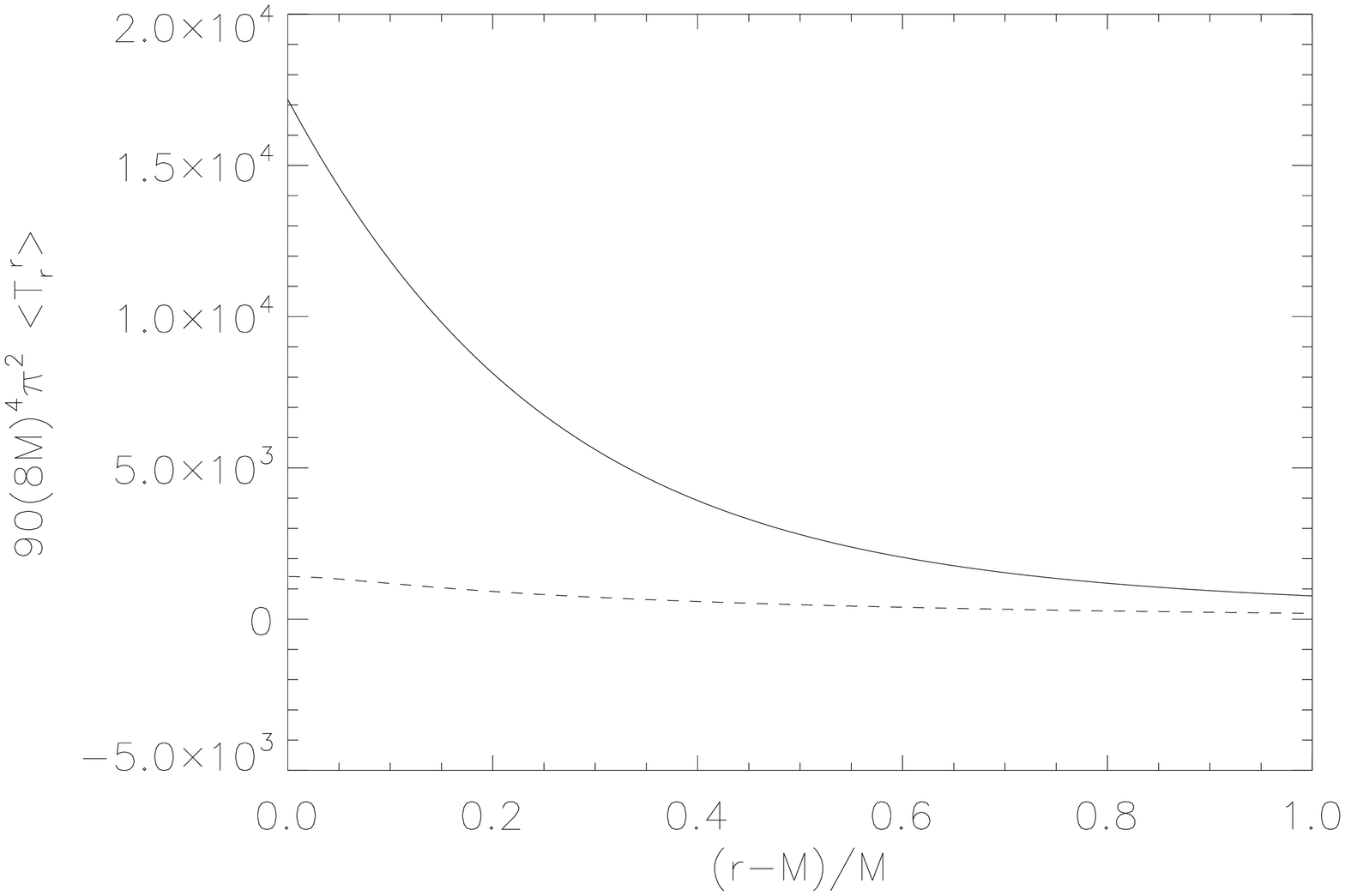}
 \vskip -.2in \caption{The
expectation value $\langle T^r_{\ r}\rangle$ for a massless spin
$1/2$ field in the extreme Reissner-Nordstr\"{o}m geometry. The solid
line corresponds to the auxiliary field stress tensor and the dashed
line to the numerically computed exact one.} \label{fig:ERNTrr12}
\end{figure}

\begin{figure}
\vskip -0.2in \hskip -0.4in
\includegraphics[angle=90,width=3.4in]
{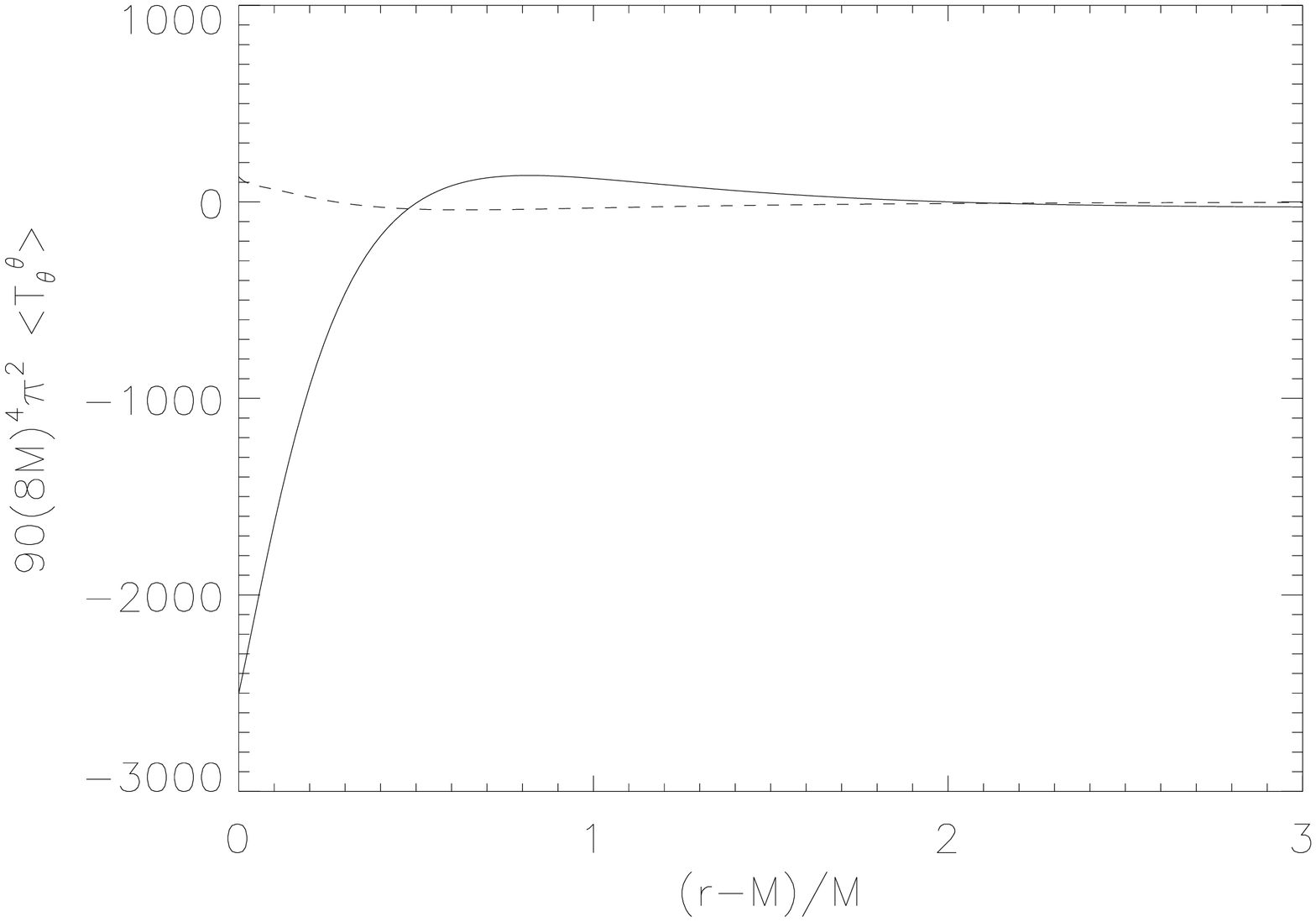}
\vskip -.2in \caption{The expectation value $\langle
T_\theta^{\ \theta}\rangle$ for a conformally invariant scalar field
in the extreme Reissner-Nordstr\"{o}m geometry. The solid line
corresponds to the auxiliary field stress tensor and the dashed line
to the numerically computed exact one.} \label{fig:ERNthth}
\end{figure}

\begin{figure}
\vskip -0.2in \hskip -0.4in
\includegraphics[angle=90,width=3.4in]
{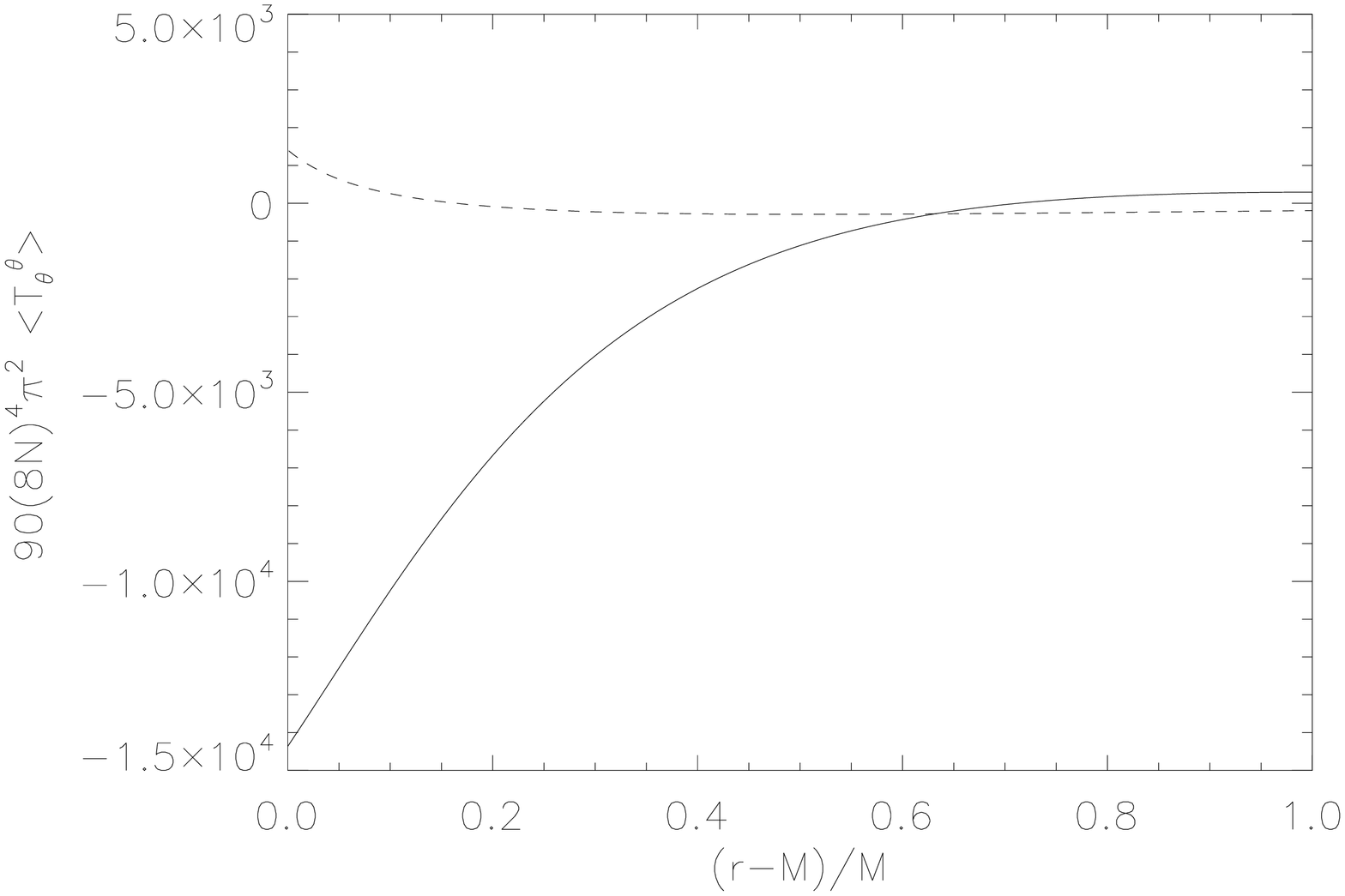}
\vskip -.2in \caption{The expectation value
$\langle T_\theta^{\ \theta}\rangle$ for a massless spin $1/2$ field
in the extreme Reissner-Nordstr\"{o}m geometry. The solid line
corresponds to the auxiliary field stress tensor and the dashed line
to the numerically computed exact one.} \label{fig:ERNthth12}
\end{figure}

\begin{figure}
\vskip -0.2in \hskip -0.4in
\includegraphics[angle=90,width=3.4in]
{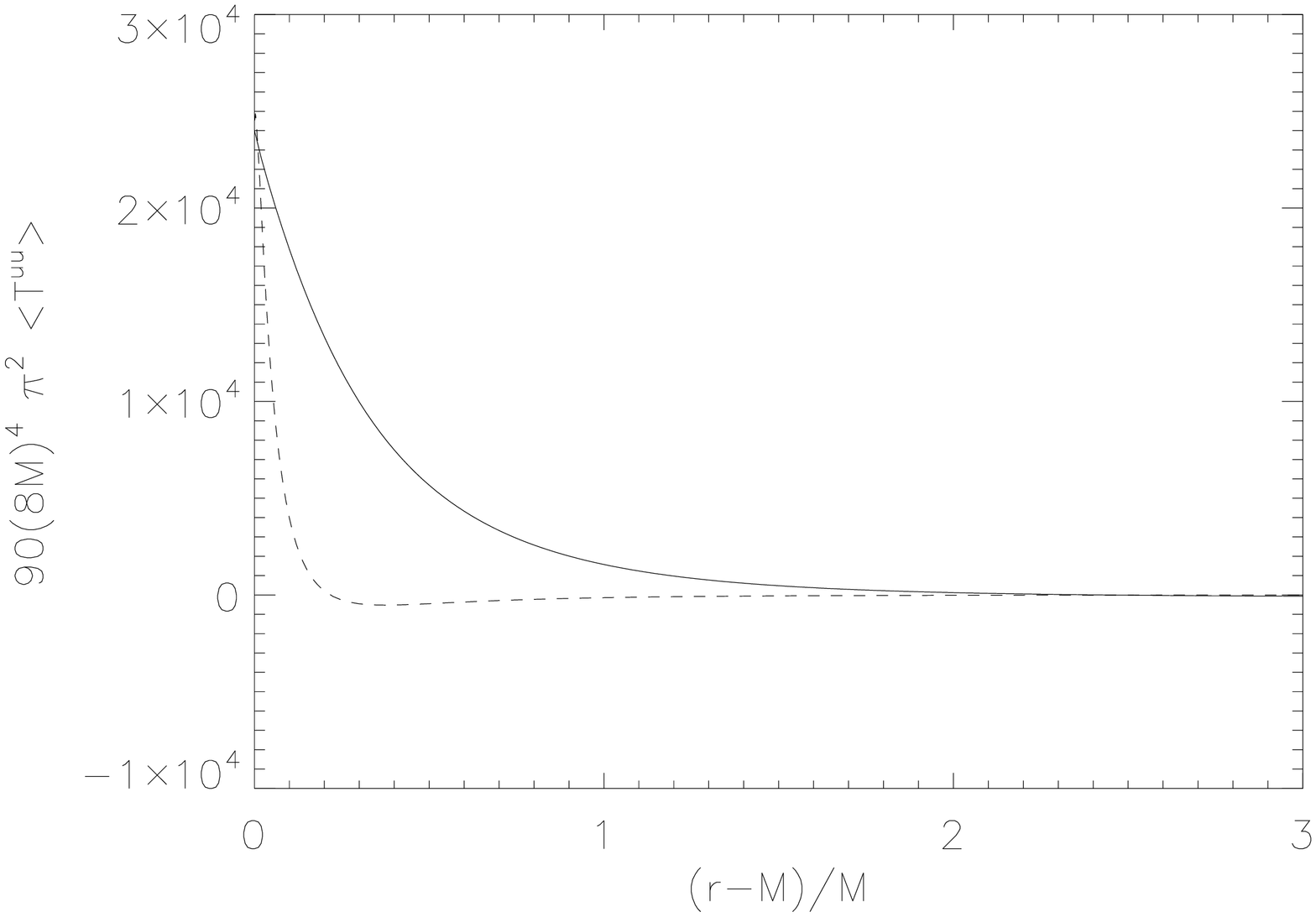}
\vskip -.2in \caption{The expectation value $\langle
T^{uu} \rangle$ for a conformally invariant scalar field in the
extreme Reissner-Nordstr\"{o}m geometry. The solid line corresponds
to the auxiliary field stress tensor and the dashed line to the
numerically computed exact one.} \label{fig:ERNuu}
\end{figure}

\begin{figure}
\vskip -0.2in \hskip -0.4in
\includegraphics[angle=90,width=3.4in]
{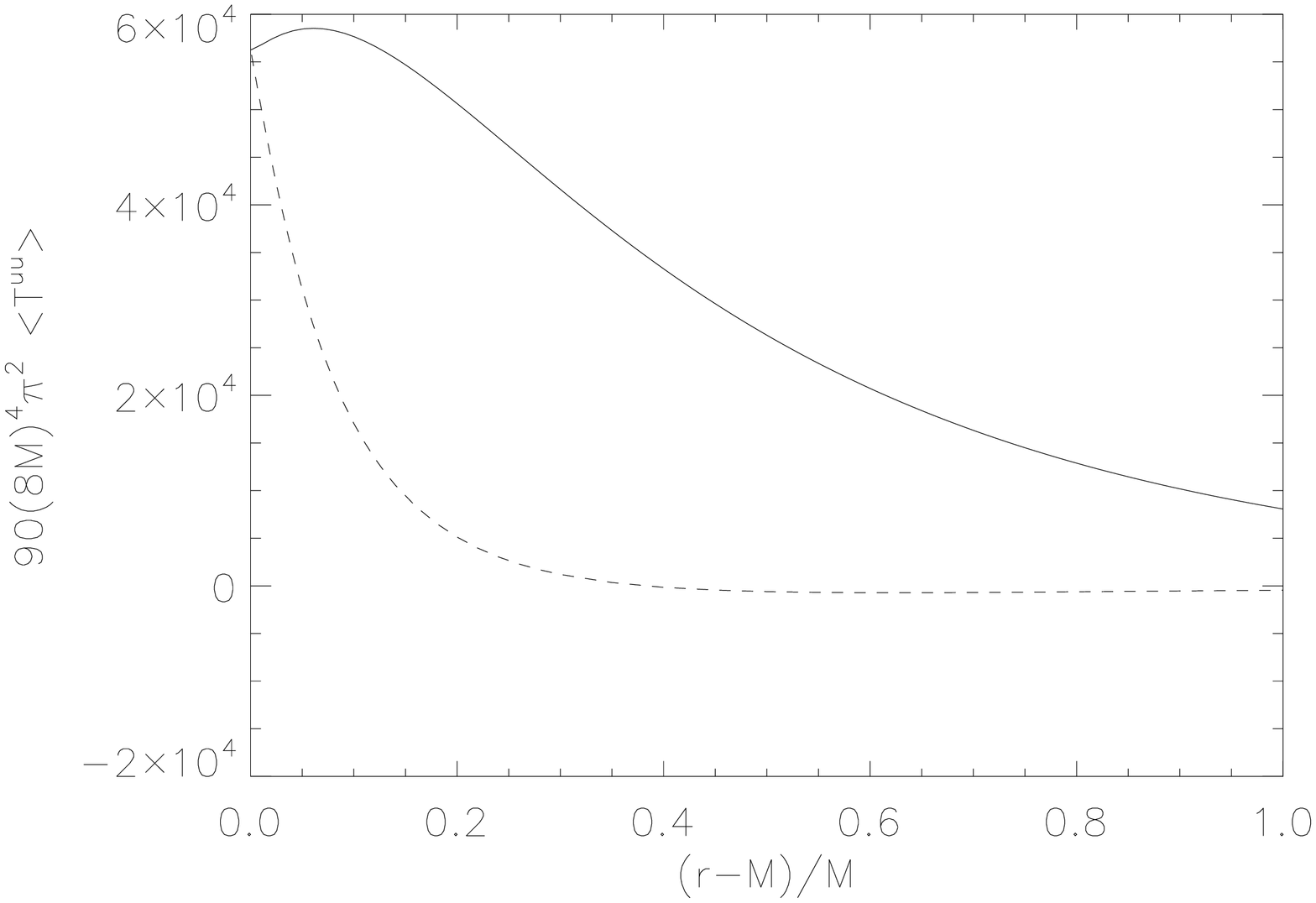}
\vskip -.2in \caption{The expectation value
$\langle T^{uu}\rangle$ for a massless spin $1/2$ field
in the extreme Reissner-Nordstr\"{o}m geometry. The solid line
corresponds to the auxiliary field stress tensor and the dashed line
to the numerically computed exact one.}
\label{fig:ERNuu12}
\end{figure}

\noindent Despite having two remaining integration constants free,
this and the other two solutions of the finiteness
conditions (\ref{ERNfinite})-(\ref{ERNuufinite}) in the ERN case
are rather restrictive. For example, it turns out that for all three
of the distinct solutions to (\ref{ERNuufinite}), the diagonal
components $T^t_{\ t}, T^r_{\ r}$, and $T^{\theta}_{\ \theta}$ on
the horizon are all fixed (for fixed  values of $b$ and $b'$) to the
specific finite values,
\begin{subequations}
\begin{eqnarray}
T^t_{\ t} = T^r_{\ r} &=&  \frac{178}{13 M^4} b - \frac{2}{M^4} b'\\  
T_\theta^{\ \theta}  & = & -\frac{178}{13 M^4} b -
\frac{2}{M^4} b' 
\end{eqnarray}
\label{ernapprox}
\end{subequations}
The correct values of these components on the ERN horizon are~\cite{AnHiTa}
\begin{equation}
T^t_{\ t} = T^r_{\ r} = T^\theta_{\ \theta} = - \frac{2}{M^4} b' \;.
\label{ernexact}
\end{equation}
The mismatch of the first term in both members of (\ref{ernapprox}), which
is traceless, accounts for the large discrepancy of the approximation 
from the numerical data apparent in Figs. 5-12. This tells us that we
are certainly lacking some term in $S_{inv}$ in our minimal approximation 
based on the anomaly, which is needed to give the correct finite coefficient of
the stress tensor on the ERN horizon. The term (\ref{Sinv}) is an example of
just such a term we have not considered.

On the other hand, the $T^{uu}$ component depends linearly on $a_0$ and $c_1$
\begin{equation}
T^{uu} = \frac{1}{585 M^4} (306856 - 9360 a_0 + 12168 c_1) b -
\frac{4484}{75 M^4} b'
\end{equation}
which are still free. Hence it is possible to adjust the value of
the $T^{uu}$ component on the horizon to any finite value and still
have one free parameter left unspecified. This last parameter can be
fixed by requiring that the behavior of all components of the stress
tensor vanish as $r\rightarrow \infty$, consistent with a zero
temperature state in an asymptotically flat spacetime.

A numerical code was developed which for a given value of $c_1$
chose the value of $a_0$ so that $T^{uu}$ matched the values
previously obtained~\cite{AnHiSa,CHOAG} from exact numerical
computations of the stress tensor on the horizon for the spin $0$
and spin $\frac{1}{2}$ fields.  As in the RN case described above, this code
then solves the equations for the auxiliary fields $\phi$ and $\psi$
and computes the analytic approximation (\ref{Tanom}) for the stress
tensor for various values of the radial coordinate $r$.

Our results for the spin $0$ and $\frac{1}{2}$ fields are shown in Figs. 5
-12.  Comparison of Eqs.\ (\ref{ernapprox}) and (\ref{ernexact})
shows that the $T^t_{\ t}$, $T^r_{\ r}$, and $T^\theta_{\ \theta}$
components are not accurate at the horizon.  The plots show
that this inaccuracy also occurs at intermediate values of $r$.
Since it is possible to choose parameters to fit the correct value
of $T^{uu}$ on the horizon, the approximation works better for that
component. Because of the more severe horizon divergences 
possible in the ERN case, the finiteness conditions
(\ref{ERNfinite})-(\ref{ERNuufinite}) eliminating them are
restrictive enough to lead to stress tensors which are not very good
approximations to the numerical results obtained by the direct
method. One possibility is to consider auxiliary field solutions 
for which there is a logarithmic divergence in $T^{uu}$ but no 
power law divergence. In this case it is possible to obtain the 
correct values of ${T_t}^t$, ${T_r}^r$, and $T^{\theta}_{\ \theta}$ 
near the horizon.  However we have found that the
approximation remains poor at intermediate and large values of $r$
and of course it also gives a divergence in $T^{uu}$ on the horizon.

A more promising approach is to add additional terms in $S_{inv}$, as for
example (\ref{Sinv}), in an attempt to find a better approximation
that is finite on the horizon. Although a modification such as this
is likely to introduce enough new parameters to allow for significant
improvement of the comparisons with the numerical results in
Figs. (\ref{fig:ERNtt}-\ref{fig:ERNuu12}), we do not pursue this possibility
here, preferring to put the minimal two field anomaly action and
stress tensor to its most stringent test. We also did not add local terms
to the effective action, such as $C_{abcd}C^{abcd}$, which certainly should
be present in general, and which give contributions to $\langle T^a_{\ b}\rangle$
of the same order as the anomaly action in regular states (where all
contributions are of order $M^{-4}$ and small for $M \gg M_{Pl}$). While
the approximation based solely on the anomaly action (\ref{Sanom})-(\ref{SEF})
itself is not very numerically accurate for ERN spacetimes, it is nevertheless
worth emphasizing that it is the first finite approximation that has been
obtained for massless quantized fields in these spacetimes, and therefore
captures some element of the exact effective action lacking in previous
approaches, even without the improvements possible with the addition
of local or Weyl invariant terms.

\section{Conclusions}

The effective action associated with the trace anomaly provides a
general algorithm for approximating the expectation value of the
stress tensor of conformal matter fields in arbitrary curved
spacetimes. It successfully classifies the leading and subleading
divergent behaviors of the quantum stress tensor components in the
vicinity of all spherically symmetric event horizons, and the conformal
properties of horizons. These behaviors follow from an analysis of 
the allowed singular behaviors of the auxiliary fields as the event 
horizon is approached, which are determined by solutions of the conformally
invariant differential operator (\ref{Deldef}). 

A numerical solution of the auxiliary field equations is necessary to
construct the stress tensor at points arbitrarily distant from the
horizon, except for the uncharged Schwarzschild case which admits a
completely analytic solution. Due to the possibility of adding
homogeneous solutions to the auxiliary field equations of motion
(\ref{EFtraces}), the generic divergent behaviors may be canceled,
and finite approximations to $\langle T^a_{\ b} \rangle$ obtained on
all charged RN event horizons, including also the extremal case of
$Q=M$.

Although it is possible to constrain the solutions of the auxiliary
field equations by the requirement that the stress tensor should be
regular on the horizon for all $Q$, we emphasize that the fine
tuning of integration constants that is necessary to achieve this
suggests that a regular stress tensor is very much the non-generic
case.  This is in accord with the well known fact that to have a
completely regular static stress-energy tensor in a static black
hole spacetime it is necessary for the field to be in a
Hartle-Hawking-Israel state~\cite{HarHaw} which is a thermal state
at the black hole temperature.  What may be less well known is
that this is also true for the zero temperature ERN black
hole~\cite{AHL} for which the Hartle-Hawking-Israel and
Boulware~\cite{Boul} states coincide.  Also the
fact of having a thermal state at a given temperature (including
zero) does not completely specify the state.  For example one can
put boundary conditions on the mode functions at some
particular value of the radial coordinate $r$ as is done when a
spherical mirror surrounding the black hole is present~\cite{Elst2,Elst3}.

The generic diverging behavior of the scalar auxiliary fields and
their associated stress tensor near the event horizon has a
geometric origin in the behavior of the Killing field of the static
geometry becoming null on the horizon, through (\ref{Kill}). Thus,
possible divergences of the stress tensor on the horizon are
perfectly consistent with the Equivalence Principle, notwithstanding
the finiteness of local curvature invariants such as $R$ or
$R_{abcd}R^{abcd}$ there.

The comparison of the anomaly induced stress tensor, viewed as an
approximation to $\langle T^a_{\ b} \rangle$ in Eq.\ (\ref{Tanom}), with the
direct evaluation of the renormalized expectation value for fields
of spin $0$ and $\frac{1}{2}$ shows that it yields a fair approximation for
states which are regular on the horizon, becoming less accurate
quantitatively both away from the horizon and in the case of the
degenerate horizon of the extreme $Q=M$ case. Although the finite
terms may be more accurately obtained by a better understanding of
the Weyl invariant non-local terms in the quantum effective action,
such as (\ref{Sinv}), the fact that the minimal anomalous terms can
give all the possible divergent or regular behaviors of the stress
tensor in the vicinities of all RN black hole event horizons allows us
to regard it as a candidate geometrical action for meaningful
backreaction calculations. Since the effective action and auxiliary
field stress tensor of the anomaly can be computed in principle in
any spacetime without regard to special symmetries, backreaction
calculations are possible with $S_{anom}$ of (\ref{Sanom}) added to
the classical Einstein-Hilbert and matter field actions. Moreover, since
the resulting system of equations can be treated by classical methods,
dynamical backreaction calculations in time dependent and even
non-spherically symmetric spacetimes undergoing gravitational
collapse, taking into account one-loop vacuum polarization and
particle creation effects, would seem to be practically feasible
by this approach for the first time.

\vskip .5cm
\centerline{{\it Acknowledgments}}
\vskip .5cm

We thank G. A. Newton for helping to develop the MathTensor
\cite{matht} code which we used to check Eqns. (\ref{Eab}) and
(\ref{Fab}) and generate those in the Appendix. We thank E. D.
Carlson, W. H. Hirsch, and B. Obermayer for the use of their
numerical data for the spin $1/2$ field. We thank P. O. Mazur and
A. Roura for related discussions. P. R. A. would like to thank the
Gravitation Group at the University of Maryland, the Department of
Theoretical Physics at the University of Valencia, and the Physics
Department at the University of Bologna for hospitality. P.R.A. also
acknowledges the Einstein Center at Hebrew University, the Forchheimer
Foundation, and the Spanish Ministerio de Educaci\'on y Ciencia for
financial support. R. V. acknowledges partial support from DOE
High Energy Physics grant E161. This research has been partially supported
by grant numbers PHY-0070981 and PHY-0556292 from the National Science
Foundation.  Numerical computations were performed on the Wake Forest University
DEAC Cluster with support from an IBM SUR grant and the Wake Forest University 
IS Department.

\vfil\eject

\appendix
\section{General Solutions of the Stress Tensor Finiteness Conditions}

In this Appendix we list the possible ways for solving the horizon
finiteness conditions in each of the three cases considered in the text.

\subsection{Schwarzschild Spacetime}

The solution of the regularity conditions in Schwarzschild spacetime
can be classified into three different groups, depending upon the
way one solves (\ref{Scond2}).

The first group is special, because one can satisfy the three
conditions (\ref{ScondA}), by specifying only two of the free constants,
namely $q=q'=2$.  This is the solution considered in \cite{MotVau}.
All of the finiteness conditions can be satisfied by specifying only
five integration constants {\it viz.},
\begin{subequations}
\begin{eqnarray}
&&q=2\\
&&q'=2\\
&& 2 bd_H=(bq\eta'+bq'\eta+b'q\eta)- 2(b + b') c_H\\
&& 2b\eta'= (bq\eta'+bq'\eta+b'q\eta) - 2(b+b')\eta\\
&& c_H= \eta \qquad {\rm or} \qquad
c_H=\frac{(bq\eta'+bq'\eta+b'q\eta)}{2b+b'}-\eta \;\;.
\end{eqnarray}
\end{subequations}

The other two groups of solutions use six instead of five constants to
satisfy the regularity conditions:
\begin{subequations}
\begin{eqnarray}
&&q=2\\
&&c_H=0\\
&&c_{\infty}=\frac{20}{9}\\
&& d_H= +\frac{q'-2}{6}\\
&& \eta'=\frac{(bq\eta'+bq'\eta+b'q\eta)}{2b}-\frac{2b'+bq'}{2b}\eta\\
&& \eta=0 \qquad {\rm or} \qquad \eta = \frac{(bq\eta'+bq'\eta+b'q\eta)}{b'+bq'}
\end{eqnarray}
\end{subequations}

\noindent
and
\begin{subequations}
\begin{eqnarray}
&& q=2-\frac{2b}{b'}(q'-2)\\
&& c_H=-\frac{2b}{b'}d_H\\
&& c_{\infty}=\frac{20}{9}-\frac{2b}{b'}\left(d_{\infty}-\frac{20}{9}\right)\\
&& d_H=-\frac{b'}{2b(2b+b')} (bq\eta'+bq'\eta+b'q\eta) +\frac{b'(q'-2)}{6(2b+b')}\\
&& \eta'=\frac{b'}{2b}\left(\frac{(bq\eta'+bq'\eta+b'q\eta) -
\eta (4b+2b'-bq')}{2b+b'-bq'}\right)\\
&& \eta=0 \qquad {\rm or} \qquad \eta =\frac{(bq\eta'+bq'\eta+b'q\eta)}{2b+b'}  \;\;.
\end{eqnarray}
\end{subequations}

\indent
In the last two groups it is assumed that $q' \neq2$. Note that with all of the above
set of constants the regularity conditions for the stress-energy tensor
are satisfied for any luminosity $L$, which is given by
\begin{equation}
L= \frac{\pi}{M^2} (bq\eta'+bq'\eta+b'q\eta)\,.
\end{equation}
This combination of parameters appears in several places in the above solutions,
so that if $L$ is set equal to zero, as it is in the Hartle-Hawking-Israel
state, considerable simplification of the relations above results.

\subsection{General RN Spacetime}

The simplest possibility for solving all five conditions
(\ref{RNfinT}) in the $0 <Q<M$ cases is the minimal one
(\ref{RNmin}).  It requires the fixing of only four integration
constants whose values are given in Eq.\ (\ref{RNmin}). The other
two distinct possibilities are:

\begin{subequations}
\begin{eqnarray}
\ell_0 & = & 0 \\
\ell_1& = & -\frac{2b}{b'}\lambda_1 \\
\lambda_1& = &\frac{6b'\lambda_0}{\epsilon (8b\lambda_0-b')}\\
a_1&=&\frac{3}{\epsilon}-\frac{2b}{b'\epsilon}[(c_1-3)\epsilon+2\lambda_0(1-3\epsilon)]\\
\lambda_0&=&\frac{3(1-\epsilon)(-6b\epsilon +(1-3\epsilon)b')}{2b(3-12\epsilon
+20\epsilon^2)}
\end{eqnarray}
\end{subequations}

\noindent and
\begin{subequations}
\begin{eqnarray}
\ell_0 & = & -\frac{2b}{b'}\lambda_0 \\
\ell_1& = & 0\\
\lambda_1& = &-\frac{6(2b\epsilon +b')\lambda_0}{\epsilon (8b\lambda_0+b')}\\
a_1&=&\frac{3}{\epsilon}+\frac{4b}{b'\epsilon}\lambda_0(1-3\epsilon)\\
\lambda_0&=&-\frac{3b'(1-\epsilon )[2b\epsilon (1-2\epsilon)+b'(1-3\epsilon)]}
{2b[2b\epsilon (8\epsilon^2 +3)+b'(3 - 12 \epsilon + 20\epsilon^2 )]}  \;\;.
\end{eqnarray}
\end{subequations}

\noindent These latter two solutions use five constants and are more
restrictive than the first minimal one in terms of four integration
constants. In particular the values of the components $T^t_{\ t},
T^r_{\ r}$ and $T_\theta^{\ \theta}$ are all fixed on the horizon in
the latter two solutions. There remains the freedom to adjust
$T^{uu}$ on the horizon, which depends upon the value of $a_0$.

\subsection{ERN Spacetime}

In addition to the solution (\ref{ERNsoln}) to the regularity conditions
(\ref{ERNfinite})-(\ref{ERNuufinite}) given in the text, two other
distinct solutions of these conditions are possible, {\it viz.}
\begin{subequations}
\begin{eqnarray}
a_{-2} & = &-\frac{2b}{b'}c_{-2}  \\
a_{-1} & = & -\frac{2b}{b'}c_{-1}\\
a_{1} &=&-\frac{60b^2(3+c_1)c_{-1}+2bb'(45+15c_1+81c_{-1}+13c_1c_{-1})+
2b'^2(45+13c_{-1})}{30bb'c_{-1}+b'^2(15+13c_{-1})}\\
c_{-2} &=& \frac{30bc_{-1}+b'(15+13c_{-1})}{8(24b+13b')}\\
c_{-1} &=& \frac{3510b^2b'+1329b'^2b+13b'^3\pm |b'(24b+13b')|
\sqrt{32400b^2+4320bb'+b'^2}}{1620b^3+3156b^2b'+1105b'^2b}  \;\;.
\end{eqnarray}
\label{ERNsqrtsols}
\end{subequations}

\indent The two solutions differ only in the sign of the square root
in the expression for $c_{-1}$. The common feature of these
solutions is that only five out seven available integration
constants are used. Nevertheless, as in the previous RN case, the
values of the components $T^t_{\ t}, T^r_{\ r}$ and $T^{\theta}_{\
\theta}$ are fixed on the horizon for all the finite solutions.
There remains only the freedom to adjust $T^{uu}$ which depends on
$a_0$ and $c_1$, both of which are still free. Below we also list
the relevant parameters corresponding to the solutions found.

For the first solution, (\ref{ERNsoln}) used in the text the values
on the horizon for a single conformal scalar field are
\begin{equation}
T^t_{\ t} = T^r_{\ r} = -\frac{2757}{90\pi^28^4M^4}
\end{equation}
and
\begin{equation}
T^{uu} = \frac{-92250 + 3072a_{0}-3994 c_{1}}{90\pi^28^4M^4} \;.
\end{equation}

For the second solution with the positive sign of the square root
in (\ref{ERNsqrtsols}), and for a single conformal scalar field,
\begin{subequations}
\begin{eqnarray}
a_{-2}& \simeq & -0.005526 \\
a_{-1} & \simeq & 1.135\\
a_{1}& \simeq &-93.83 + 6 c_{1}\\
c_{-2}&\simeq &-0.000921\\
c_{-1}& \simeq &0.1892 \;.
\end{eqnarray}
\end{subequations}
The values on the horizon are
\begin{equation}
T^t_{\ t} = -\frac{746}{90\pi^28^4M^4}\,
\end{equation}
and
\begin{equation}
T^{uu} = \frac{-465000 + 3072 a_{0}+31070
c_{1}}{90\pi^28^4M^4} \;\;.
\end{equation}

For the second solution with the negative sign of the square root
in (\ref{ERNsqrtsols}), and for a single conformal scalar field,
\begin{subequations}
\begin{eqnarray}
a_{-2}& \simeq & -3.274 \\
a_{-1} & \simeq & -18.90\\
a_{1}& \simeq &-5.306 + 6c_{1}\\
c_{-2}& \simeq &-0.5457\\
c_{-1}& \simeq &-3.151 \;\;.
\end{eqnarray}
\end{subequations}
The values on the horizon are
\begin{equation}
T^t_{\phantom{t}t} = \frac{10320}{90\pi^28^4M^4}
\end{equation}
and
\begin{equation}
T^{uu}=\frac{25700 + 3072 a_{0}-22980 c_{1}}{90\pi^28^4M^4} \; \;.
\end{equation}

\end{document}